\documentclass[preprint2,longabstract]{aastex}

\newcommand{\etal}{et~al.} 
\newcommand{\ionhy}{H{\sc ii} }
\newcommand{\UCHII}{UCH{\sc ii} }

\newcommand{\kms}{$\mbox{km~s}^{-1}$ }
\newcommand{\kmsns}{$\mbox{km~s}^{-1}$}

\shorttitle{Statistical properties of 12.2~GHz methanol masers}
\shortauthors{Breen et al.}

\begin{document}


\title{Statistical properties of 12.2~GHz methanol masers associated with a complete sample of 6.7~GHz methanol masers}


%
%
%
%
%
%
%
%
%
%
%
%
%
%
%
%

\author{S. L. Breen,$^{1,2}$ S. P. Ellingsen,$^{2}$ J. L. Caswell,$^{1}$ J. A. Green,$^{1}$ G. A. Fuller,$^{3}$ M. A. Voronkov,$^{1}$ L. J. Quinn,$^3$ A. Avison$^3$\\
$^1$ CSIRO Astronomy and Space Science, PO Box 76, Epping, NSW 1710, Australia\\
$^2$ School of Mathematics and Physics, University of Tasmania, Private Bag 37, Hobart, Tasmania 7001, Australia\\
$^3$ Jodrell Bank Centre for Astrophysics, Alan Turing Building, School of Physics and Astronomy, University of Manchester, Manchester M13 9PL}

\email{Shari.Breen@csiro.au}

\begin{abstract}

We present definitive detection statistics for 12.2~GHz methanol masers towards a complete sample of 6.7~GHz methanol masers detected in the Methanol Multibeam survey south of declination --20$^{\circ}$. In total, we detect 250 12.2~GHz methanol masers towards 580 6.7~GHz methanol masers. This equates to a detection rate of 43.1 \%, which is lower than that of previous significant searches of comparable sensitivity.  Both the velocity ranges and the flux densities of the target 6.7~GHz sources surpass that of their 12.2 GHz companion in almost all cases. 80 \% of the detected 12.2 GHz methanol maser peaks are coincident in velocity with the 6.7 GHz maser peak. Our data support an evolutionary scenario whereby the 12.2~GHz sources are associated with a somewhat later evolutionary stage than the 6.7~GHz sources devoid of this transition. Furthermore, we find that the 6.7~GHz and 12.2~GHz methanol sources increase in luminosity as they evolve. In addition to this, evidence for an increase in velocity range with evolution is presented. This implies that it is not only the luminosity, but also the volume of gas conducive to the different maser transitions, that increases as the sources evolve. Comparison with GLIMPSE mid-infrared sources has revealed a coincidence rate between the locations of the 6.7~GHz methanol masers and GLIMPSE point sources similar to that achieved in previous studies. Overall, the properties of the GLIMPSE sources with and without 12.2~GHz counterparts are similar. There is a higher 12.2~GHz detection rate towards those 6.7~GHz methanol masers that are coincident with extended green objects.

\end{abstract}


\section{Introduction}

Masers of water, methanol and hydroxyl are considered excellent probes of high-mass star formation regions. They are common, bright and are observed during the phase where the forming stars are still deeply embedded in their natal material. The different maser species require slightly different physical conditions for their production and survival \citep[e.g.][]{Cragg02}, which have led to proposals which base high-mass star formation evolutionary schemes on the presence or absence of the different maser transitions \citep[e.g.][]{Ellingsen07,Breen10a,Fontani10}. 

Previous searches for 12.2~GHz methanol masers have typically targeted \UCHII regions \citep[e.g.][]{Batrla87,Norris1987} or 6.7~GHz methanol masers that were detected towards \UCHII regions, OH masers or {\it IRAS} selected regions \citep[e.g.][]{Gay,Caswell95b,Blas04,Breen10a}. OH masers have been noted on many occasions as being associated with a generally later stage of star formation than other maser species \citep[e.g.][]{FC89,Cas97}. Furthermore, those {\em IRAS} selected searches were targeted towards sources that had colours similar to \UCHII regions based on the \citet{WoodChurch89} criteria. Each of the aforementioned 12.2~GHz methanol maser searches have therefore been biased towards sources of a generally later stage of high-mass star formation since we know that many 6.7~GHz methanol masers are devoid of one or both of associated OH masers \citep{Cas96} and \UCHII regions \citep{Walsh98}. 

\citet{Breen10a} used their 12.2~GHz methanol maser observations (together with complementary data) to investigate the relative evolutionary stages of the high-mass star formation that the 6.7 and 12.2~GHz methanol maser transitions were tracing. From this analysis \citet{Breen10a} concluded that 12.2~GHz methanol masers were present towards the second half of the 6.7~GHz methanol maser lifetime. They additionally found that both methanol transitions appeared to increase in luminosity as the sources evolved, although at a slower rate and by a lesser amount in the case of the 12.2~GHz sources. Although comprising a relatively small and biased sample of sources, these results are very encouraging in the overall goal to use the common maser species as evolutionary tracers of high-mass star formation.

The idea that the presence, absence or intensity of certain properties associated with high-mass star formation regions can give an indication on the relative evolutionary stage of star formation is not a new one. In particular, from soon after the discovery of interstellar masers, it has been suggested that these masers are associated with a distinct evolutionary phase of star formation \citep[e.g.][]{LBH75,GD77} and since this time, many maser studies have found evidence that this is the case \citep[e.g.][]{FC89,Cas97,Walsh98,Ellingsen06,Breen10a,Fontani10}. Evolutionary scenarios based on combinations of specific infrared, submillimetre and millimetre radio continuum properties are also popular and show very promising results \citep[e.g.][]{Cyg08,Chambers09,Rath10}. Thermal lines have also proven to be excellent evolutionary probes, especially since, in addition to the presence or absence of specific transitions, line properties can be used to further pinpoint the source's evolution \citep[e.g.][]{Purcell06p,Purcell09,Longmore07}. The implied evolutionary stage that is deduced by these different methods appear to be in excellent agreement. For example, \citet{Longmore07} found that the linewidths of the detected ammonia emission increased with evolution. Comparison between ammonia linewidths and methanol maser luminosities by \citet{Wu10} show that the lower luminosity 6.7~GHz methanol masers are associated with smaller ammonia linewidths and therefore correspond to the younger sources. This corroborates the findings of \citet{Breen10a}, that the 6.7~GHz methanol masers increase in luminosity as they evolve.

In this paper we present the statistical properties of the 12.2~GHz methanol maser emission detected towards a complete sample of 6.7~GHz methanol masers. In particular, we test the conclusions drawn in \citet{Breen10a} on a much larger and less biased sample of sources. Throughout this work we place a particular emphasis on investigating the properties of the different maser species at different evolutionary stages. The 6.7~GHz methanol maser targets comprise a complete sample of almost 600 methanol masers detected in the Methanol Multibeam Survey (described in Section~\ref{sect:mmb_intro}). In a subsequent publication we will be presenting the catalog of detected 12.2~GHz methanol masers, together with discussion of individual source properties that can be best addressed by looking at their specific data. This publication order is necessary since many of our target sources are yet to be published.

\subsection{The Methanol Multibeam (MMB) Survey }\label{sect:mmb_intro}

The Methanol Multibeam (MMB) Survey for 6.7~GHz methanol masers is the largest complete survey for 6.7~GHz methanol masers ever undertaken. The complete survey aims to cover the full Galactic Plane within $b$ = $\pm$2$^{\circ}$. A thorough description of the survey parameters and techniques can be found in \citet{Green09}. 

The southern hemisphere component of the MMB survey has detected nearly 1000 6.7~GHz methanol masers (to a detection limit of $\sim$0.7~Jy), the majority of which have now been followed up with interferometric observations in order to obtain accurate source positions. Survey results from the Galactic longitude range 345$^{\circ}$ to 20$^{\circ}$ have already been released \citep{CasMMB10,GreenMMB10} and results from the range 186 to 345$^{\circ}$ are forthcoming.

\section{Observations and data reduction}

\subsection{6.7~GHz MMB maser targets}

12.2~GHz follow up observations towards 593 6.7~GHz methanol masers detected in the MMB survey were carried out with the Parkes radio telescope during 2008 June and December. The sources searched correspond to all of the MMB detections that lie south of declination --20 degrees. This declination cutoff was imposed for several reasons, foremost of which was that at the time of the 12.2~GHz observations, accurate positions were chiefly limited to those sources that fell in this range. Since the positional uncertainties of sources detected in the initial survey were comparable to the half power beam width (HPBW) of the Parkes radio telescope at 12.2~GHz, follow up observations towards the sources that lie north of declination --20 degrees have been delayed until accurate positions are known.

As part of the MMB observing strategy, once accurate positions for the survey detections were obtained (chiefly using the Australia Telescope Compact Array, achieving positional accuracies of $\sim$0.4 arcsec), pointed observations of each source were completed with Parkes \citep{Green09}. The resulting spectra have typical rms noise levels of $\sim$0.07~Jy and spectral resolutions of 0.11~\kmsns. It is with these spectra that we compare our follow-up 12.2~GHz observations.

The 6.7~GHz methanol maser properties (velocity ranges and peak, peak flux density and integrated flux density) of the sources targeted for 12.2~GHz emission have been determined for the present analysis independently of those listed in the survey publications that have already been released \citep{CasMMB10,GreenMMB10}. However large discrepancies between values are not expected. For some sources in regions of confusion, the interferometric data had to be inspected to determine which spectral features were associated with which position. 

\subsection{12.2~GHz observations}

The search for 12.2~GHz methanol masers were made with the Ku-band receiver on the Parkes 64 m radio telescope, using an identical setup to that described in \citet{Breen10a}. This receiver detected two orthogonal linear polarizations and had typical system equivalent flux densities of 205 and 225 Jy for the respective polarizations throughout the observations in 2008 June, and were slightly higher at 220 and 240~Jy in 2008 December. The Parkes multibeam correlator was configured to record 8192 channels over 16 MHz for each of the recorded linearly polarized signals. This configuration yielded a usable velocity coverage of $\sim$290~\kms and a spectral resolution of 0.08 \kmsns, after Hanning smoothing of the data. The Parkes radio telescope has rms pointing errors of $\sim$10 arcsec and at 12.2~GHz the telescope has a HPBW of 1.9 arcmin. For two sources, no usable data was obtained in either the 2008 June or December observations. Instead, data from subsequent observations made during 2010 March are used.

The observation strategy was crafted such that the majority of the target 6.7~GHz methanol masers had a total of 10 minutes of integration time at 12.2~GHz over the two main observing epochs. This integration time yields typical rms noise limits of $\sim$0.11~Jy (corresponding to a 5-$\sigma$ detection limit of 0.55~Jy). A mixture of 10 and 5 minute integrations were taken at the  
first epoch. Sources observed for only 5 minutes (during the first epoch) which did not yield a  
detection were typically observed for an additional 5 minutes during the second epoch. Confident detections were not routinely repeated, regardless of their integration times. For 55 of the target 6.7~GHz methanol masers, the 12.2~GHz methanol maser data used in this analysis is the same as published in \citet{Breen10a}.


All sources were observed at a fixed frequency of 12178~MHz
(i.e. with no Doppler tracking), which alleviated the requirement for
a unique reference observation to be made for each of the source
positions. Instead, all the spectra collected each day were combined
to form a sensitive reference bandpass by taking the median value of
each channel. Since the target sources were spread over a large Galactic longitude range, sources were detected at different frequencies within the bandpass and this combined with the narrow nature of the sources means that this technique works very well, even in the case where a high number of strong detections are expected.  The noise contribution to the
quotient spectrum from such a reference is negligible.

Data were reduced using ASAP (ATNF Spectral Analysis Package). Alignment of velocity channels was carried out during processing. The adopted rest frequency was 12.178597 GHz \citep{Muller04}. Absolute flux density calibration was achieved by observing PKS B1934--638 each day which has an assumed flux density of 1.825 Jy at 12178-MHz \citep{Sault03}. Calibration observations were very stable during each session.

Sources that were observed at both epochs were first reduced individually and searched for detections and then the data were averaged and searched once more in order to pick up any stable but weak emission. Only three sources were identified in this manner (out of 208 sources that were observed at both epochs with no detection at either epoch). Given the luxury of knowing the velocity of the 6.7~GHz maser emission, we have been able to readily and reliably identify sources that are much weaker than the 5-$\sigma$ detection limits. A number of weak sources can confidently been regarded as genuine at 3- to 4-$\sigma$ levels. Marginal detections were considered along with their associated 6.7~GHz spectrum, and were assessed on an individual case basis.

\section{Results}\label{sect:results}

Observations conducted with the Parkes radio telescope over two epochs resulted in the detection of 250 12.2~GHz methanol masers towards 580 6.7~GHz methanol masers detected in the MMB survey, a detection rate of 43~\%. A search of the literature indicates that 148 of the 250 12.2~GHz methanol masers that we detect are new discoveries. Of the 102 12.2~GHz sources that have been previously discovered, the majority of these were first detected by \citet{Caswell95b} or \citet{Breen10a}, but others were first detected by: \citet{Gay};  \citet{Koo88}; \citet{Catarzi93}; \citet{Kemball88}; \citet{Batrla87}; \citet{Cas93}; \citet{Norris1987}; or \citet{MacLeod93}. The observations targeted a total of 593 sources south of our declination cutoff of --20 degrees but 13 sources have been removed from the detection statistics, chiefly due to our inability to distinguish them from nearby sources (a consequence of our single dish observations). Ten of these 13 sources are located close to the Galactic centre.

%


\section{Discussion}~\label{sect:dis}

\subsection{Detection statistics}\label{sect:det}


Our detection rate of 43~\% is significantly lower than previous searches of comparable sensitivity. For example, the detection rates found in the large, sensitive surveys of \citet{Caswell95b} and \citet{Breen10a} were 55~\% and 60~\% respectively. In order to determine the reasons for this apparent detection rate discrepancy, we must first examine the nature of the target sources used in previous searches. \citet{Caswell95b} conducted their search towards 6.7~GHz methanol masers that had been detected towards OH maser targets, while the observations of \citet{Breen10a} targeted 6.7~GHz methanol masers that had been detected towards {\em IRAS} selected regions \citep{Walsh97,Walsh98}. \citet{Blas04} searched 261 6.7~GHz methanol masers and achieved a detection rate of 19~\%. Their survey was of much lower sensitivity (accounting for the lower detection rate) and was also primarily directed towards 6.7~GHz methanol masers detected towards {\em IRAS} selected targets. \citet{Gay} carried out a search for 12.2~GHz methanol masers towards infrared selected 6.7~GHz targets with a similarly low sensitivity, and achieved a detection rate of 27~\%. 

Given the high sensitivity and large scale of our observations, we can only directly compare our results with those of \citet{Caswell95b} and \citet{Breen10a}; however, we are able to make some comparison with less sensitive searches by taking detection limits into account. For example, \citet{Gay} had a sensitivity limit of $\sim$10~Jy, if our detection limit was at this level our detection rate would be slightly lower than 10~\% (compared to their detection rate of 27~\%). If our detection limit had been 4~Jy, like that of the search by \citet{Blas04}, our detection rate would have dropped to 16~\% (compared to their detection rate of 19~\%). It is therefore evident that even by taking detection limits of other surveys into account, our detection rate is lower than that of previous searches.

The large searches of \citet{Caswell95b} and \citet{Breen10a} were targeted towards 6.7~GHz methanol masers that were detected in observations of comparable sensitivity (\citet{Caswell95b} and \citet{Walsh97} respectively) to that of our target 6.7~GHz sources. This means that the difference in 12.2~GHz detection statistics cannot be accounted for by the sensitivity in the searches of the target 6.7~GHz sources. The marked drop in detection rate from our search compared to similar searches can therefore most easily be accounted for by the absence of 6.7~GHz methanol maser target biases that were present in many previous searches. Our 6.7~GHz MMB targets were detected in a sensitive and unbiased manner. In contrast, all of the previous searches were subject to different levels of evolutionary biases, since both {\em IRAS} infrared colour and OH targeted searches would miss the majority of very young methanol masers. Add to this the fact that 12.2~GHz methanol masers are expected to be present at a time somewhat later than the onset of the 6.7~GHz emission \citep[e.g.][]{Breen10a}, and it is not surprising that our detection rate would be lower than such searches. \citet{Blas04} suggested that 12.2~GHz detection rates depended mainly on sensitivity; however, our search shows that the role of biases is perhaps an equally important factor.

\subsection{Galactic distribution of methanol masers}\label{sect:galdist}

\citet{Pestalozzi05} compiled a catalog of the 6.7~GHz methanol masers that were known prior to 2004. A total of 519 6.7~GHz sources were listed throughout the Galaxy and were detected in both targeted and blind searches (not all of which had been searched for 12.2~GHz counterparts). The distribution of the 6.7~GHz sources was found by the authors to be non-random throughout the Galactic plane with a particularly high concentration between 20 to 50$^{\circ}$ longitude and 330 to 340$^{\circ}$ longitude. \citet{Pestalozzi05} also found that the latitude distribution of sources across the Galactic plane had a FWHM of 0.5$^{\circ}$ and commented that this could be attributed to the fact that blind searches had been limited to low Galactic latitudes. However, a comparable narrow latitude distribution  
is seen in the MMB  \citep{CasMMB10,GreenMMB10} suggesting it is inherent to  
the population rather than a product of observational bias.

The Galactic distribution of the 6.7~GHz methanol masers detected in the MMB survey is beyond the scope of this study and we only include preliminary findings which allow us to comment on the distribution of the associated 12.2~GHz sources. The sample of MMB 6.7~GHz methanol masers south of --20$^{\circ}$ (corresponding to longitude ranges of 290$^{\circ}$ through 360$^{\circ}$ to 10$^{\circ}$) does not include the densely populated 20$^{\circ}$ to 50$^{\circ}$ longitude range that is covered by the \citet{Pestalozzi05} compilation, but encompasses a larger group of sources. In the overlapping ranges, the distribution is similar with a significant density of sources in the 330 to 340$^{\circ}$ longitude range, and a distinct lack of sources between longitudes of 250 to 290$^{\circ}$.

\begin{figure}
\includegraphics[angle=270,scale=.40]{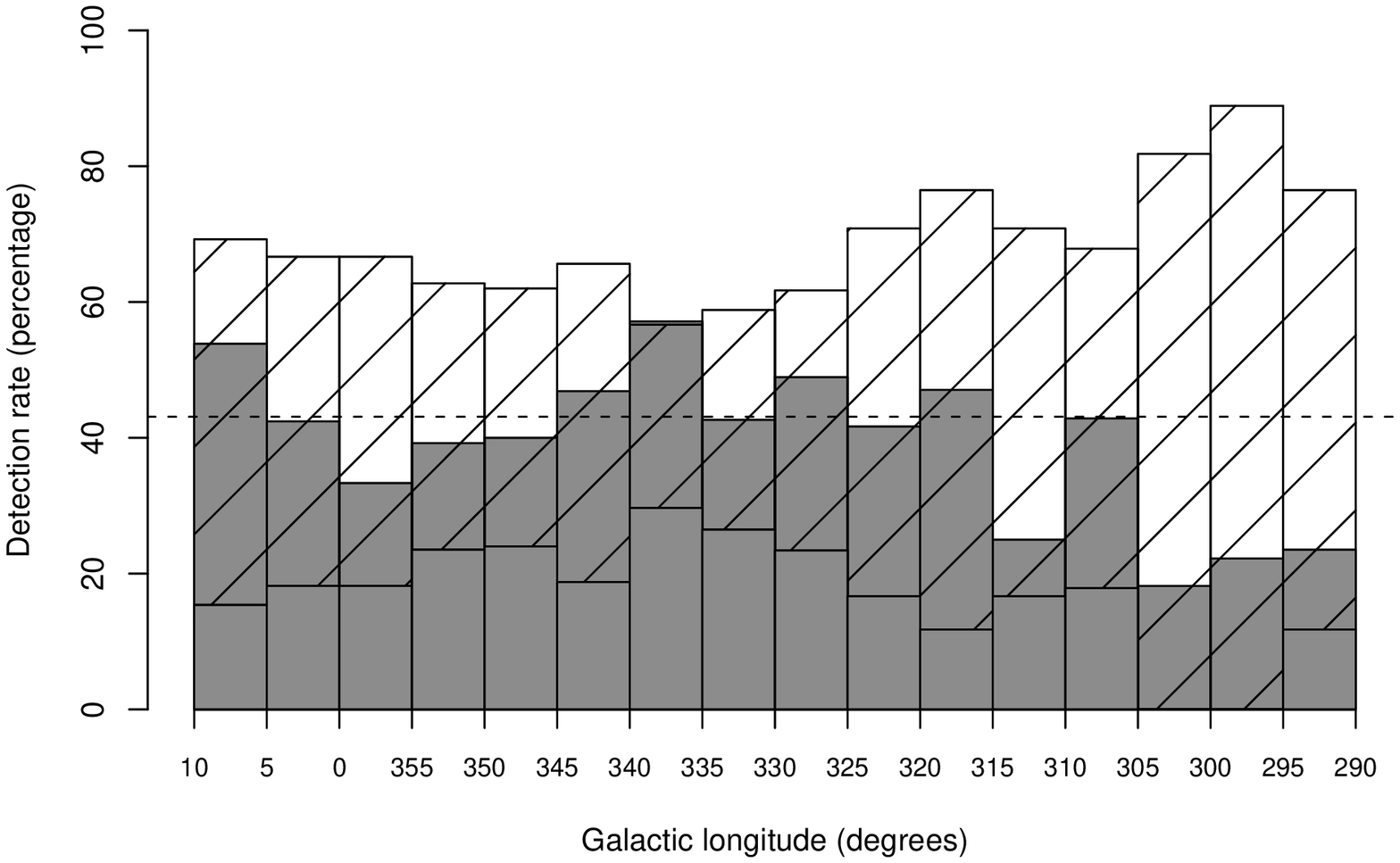}
\caption{Plot of the 12.2~GHz detection rates as a function of Galactic longitude ranges. The gray bars correspond to the actual detection rates in each longitude range and the dashed horizontal line represents the average detection rate of 12.2~GHz sources. For the detection rate to be significantly different that the mean detection rate, they would have to fall onside the diagonally shaded bars ({\em p}-value=0.05).}
\label{fig:hist_long_detections}
\end{figure}

A comparison of the detection rate of 12.2~GHz methanol masers split into 5$^{\circ}$ longitude bins shows that the detection rate is somewhat dependent on Galactic longitude range (shown in Fig.~\ref{fig:hist_long_detections}). Sources with longitudes less than 290$^{\circ}$ or more than 10$^{\circ}$ were omitted from this investigation as these 5$^{\circ}$ bins contained fewer than 5 sources, making the statistics meaningless. 

In Fig.~\ref{fig:hist_long_detections}, the the diagonally shaded bars show the range of detection rates in each longitude range that would not be significantly different from the mean detection rate. Therefore, in order for the detection rate to be statistically significantly different from the mean, it must fall outside of these diagonally shaded bars. For the 335 to 340$^{\circ}$ range, the detection rate differs from the mean detection rate significantly ({\em p}-value=0.04 from a chi-squared test) and shows a relative overabundance of 12.2~GHz sources. This may indicate that a burst of high-mass star formation occurred in this region at a time which now causes an abnormally high fraction of sources to be at similar evolutionary stages. 

An apparent relative under-abundance of 12.2~GHz detections at longitudes between 270 to 305$^{\circ}$ was further investigated and compared to a region of the same size, 315 to 350$^{\circ}$. The detection rate in the first region is 25~\% and in the second the detection rate almost doubles to a rate of 48~\% and neither rate differs significantly from the mean. However, the detection rates in these regions are statistically different from one another (tested by calculating the expected number of detections in 270 to 305$^{\circ}$ using the detection rate of the 315 to 350$^{\circ}$ region and carrying out a chi-squared test). In the 270 to 305$^{\circ}$ region, there is also a distinctly lower density of 6.7 GHz methanol masers than in other parts of the Galaxy. In the Large Magellanic Cloud (LMC), the heavy metal and complex molecule abundances are a likely explanation for the lack of 6.7~GHz methanol masers, which are underabundant by a factor of $\sim$4-5 compared to our Galaxy after accounting for the different star formation rates \citep{GreenLMC}.

As evident in \citet{Pestalozzi05} and inspection of the distribution of the MMB sources, relatively few 6.7~GHz methanol masers reside in the longitude range 270 to 305$^{\circ}$. Considering the explanation of the underabundance of methanol maser sources located in the LMC, it is possible that the underabundance of 6.7~GHz methanol masers in the 270 to 305$^{\circ}$ region could be attributed to low metallicity, which if it is the cause, may also be responsible for influencing the 12.2~GHz detection rate (which is lower than expected). \citet{Ellingsen2010} carried out a search for 12.2~GHz methanol maser emission towards the four known 6.7~GHz methanol masers in the LMC, detecting one source. Comparison of the 12.2~GHz detection rate in the LMC compared to that in the Galactic plane is difficult since the sample size in the LMC is small and achieving sensitivities comparable to sources located within our Galaxy is difficult. \citet{Ellingsen2010} comment that observations of two orders of magnitude greater sensitivity would be required to confidently determine if the sources in the LMC were under-luminous at 12.2~GHz compared to the Galactic sources.

\subsubsection{Clustering of sources exhibiting the same methanol transitions}\label{sect:clust}

The locations of the two methanol maser transitions were further inspected on a longitude-latitude plot. This plot showed an apparent clustering of the sources that exhibited masers at both transitions, and likewise for the sources exhibiting only 6.7~GHz emission. For all sources harboring 12.2~GHz emission, the nearest neighbor spatially has been examined to determine if there is a higher proportion of sources with 12.2~GHz emission near other 12.2~GHz sources than should be expected from a random distribution. Using this method we find that approximately 50~\% of 12.2~GHz sources have another source with 12.2~GHz emission as its nearest neighbor (on the longitude-latitude plot) compared to the expected proportion (based on detection statistics) of 43~\%. The difference between these values is not quite statistically significant.

The apparent clustering was further tested by investigating the nearest neighbor of each of the 12.2~GHz methanol masers on a longitude-velocity plot. This was done by determining the linear separation between each of the 12.2~GHz methanol sources and all other sources on the plot. The difference in longitude between source was multiplied by two, making the ranges on each of the axes approximately equivalent. We find that for 51~\% of 12.2~GHz methanol sources, the nearest source also exhibits 12.2~GHz emission which is statistically significantly higher that what would be expected from the 12.2~GHz detection rate ({\em p}-value of 0.05). Since we have limited our investigations to two dimensions in each case it is difficult to draw meaningful conclusions, but they do suggest that further, more sophisticated three dimensional analysis is warranted once distances to the sources are known. 

Clustering of sources showing 12.2~GHz methanol masers, combined with the non-uniform detection statistics along the Galactic plane (as discussed in Section~\ref{sect:galdist}), may imply that localised bursts of high-mass star formation have resulted in a number of nearby sources being at similar evolutionary stages. Another explanation may be that localised metallicity gradients influence the 12.2~GHz methanol maser detection statistics.

\subsection{Comparison of flux densities and velocity ranges}\label{sect:comp}

Previous searches have shown that 12.2~GHz methanol maser emission is generally weaker than their 6.7~GHz counterparts. A large sample of spectra demonstrating this is presented by \citet{Cas+95} and illustrate many of the points made here. We, too, find that 12.2~GHz methanol masers are, in general, weaker than their 6.7~GHz counterparts and that our 12.2~GHz detections are primarily at the stronger 6.7~GHz sites. Table~\ref{tab:fv_comp} presents both the average and median peak flux densities of the 6.7~GHz methanol masers (of the full sample as well as in the categories of with and without 12.2~GHz counterparts) and the 12.2~GHz sources that we detect.  The average peak to peak ratio of 6.7~GHz to 12.2~GHz methanol masers (for detected sources) is 18:1 and the median ratio is 4.3:1. Comparison of integrated flux densities shows an even greater difference in 6.7 and 12.2~GHz methanol maser strength; the average 6.7 to 12.2~GHz integrated flux density is 41:1 and have a median ratio of 10:1. 


\begin{table}
\begin{center}
  \caption{Average and median peak flux densities of both the 6.7~GHz and 12.2~GHz methanol masers.}
  \begin{tabular}{lcccc}\hline\label{tab:fv_comp}
 {\bf Maser} & \multicolumn{2}{c}{\bf Peak flux density (Jy)}\\
{\bf Category} &         {\bf Average}   &  {\bf Median} \\ \hline
All 6.7~GHz sources & 48 &5\\
6.7~GHz with 12.2~GHz & 106 & 15\\
6.7~GHz no 12.2~GHz & 6.8 & 2.5\\ \\
All 12.2~GHz sources & 23 & 2.0\\
\hline
	\end{tabular}
\end{center}

\end{table}

We find that the average velocity range of the 580 target 6.7~GHz methanol masers is 6.9 \kmsns, and they have a median range of 5.7 \kmsns. There are only 8 instances where the 6.7~GHz velocity range surpasses 20 \kmsns, and in fact, this number only increases to 41 for velocity ranges greater than 15 \kmsns. The detected 12.2~GHz methanol maser sources have smaller velocity ranges; only 22 sources exceed a velocity range of 10~\kmsns. The average velocity range of the 12.2~GHz detections is  3.2 \kmsns, and the median range is 1.7~\kmsns. Interestingly, there are very few instances (14 of 250) where a 12.2~GHz methanol counterpart is detected toward a 6.7~GHz methanol maser with a velocity range less than 2~\kmsns, even though 112 6.7~GHz methanol masers fall into this category. Therefore, we find a much lower 12.2~GHz detection rate (12.5~\%) towards 6.7~GHz methanol masers with narrow velocity ranges (not greater than 2 \kmsns). Not surprisingly, these sources have an average peak flux density that is lower than that of the sources with the larger velocity ranges.

Since the 12.2~GHz methanol masers are generally weaker than the target 6.7~GHz methanol masers, one might expect that sensitivity limitations have prevented us from detecting 12.2~GHz emission towards the weaker 6.7~GHz methanol masers. Alternatively, this effect may be limited to only the weaker features of some 6.7~GHz methanol masers. If these weaker 6.7~GHz features were located towards the minimum and/or maximum velocities of the emission, then this might affect the recorded velocity ranges of the 12.2~GHz sources.  However investigations presented in Sections~\ref{sect:peak_vel} and~\ref{sect:individualf} show that, in the majority of cases, sensitivity has not limited us from gaining a complete picture of the two methanol maser transitions. Furthermore, given that the integrated flux density ratios of the two transitions are vastly different, this suggests that even in the cases where we detect both methanol transitions there is a larger gas volume conducive to the 6.7~GHz transition than the 12.2~GHz transition.

\subsubsection{6.7 and 12.2~GHz methanol maser peak velocities}\label{sect:peak_vel}

Approximately 80~\% of the 12.2~GHz peak velocities correspond to the peak velocity of the associated 6.7~GHz methanol maser. In all but one case, the peak velocity of the remaining 20~\% of 12.2~GHz sources corresponds to the velocity of another spectral feature of the 6.7~GHz source. In some of these cases, it is the weakest feature in the 6.7~GHz maser spectrum that shows emission at 12.2~GHz, but more commonly we find the 12.2~GHz emission at the velocity of the second strongest 6.7~GHz methanol maser feature. 

There are very few weak 12.2~GHz sources with stable peak flux densities (i.e. those that show variations of less than the noise level of the observations). 236 of the 6.7~GHz methanol masers were searched at both epochs and 28 of these resulted in a detection at only one of these, while the remaining 208 sources showed no identifiable emission at either epoch. When the data for these sources were averaged over the two epochs, only three additional sources were identified. This shows that multiple epoch searches have a much higher merit than one-time searches, even when the observations are of a higher sensitivity.

\begin{figure}
\includegraphics[angle=270,scale=.40]{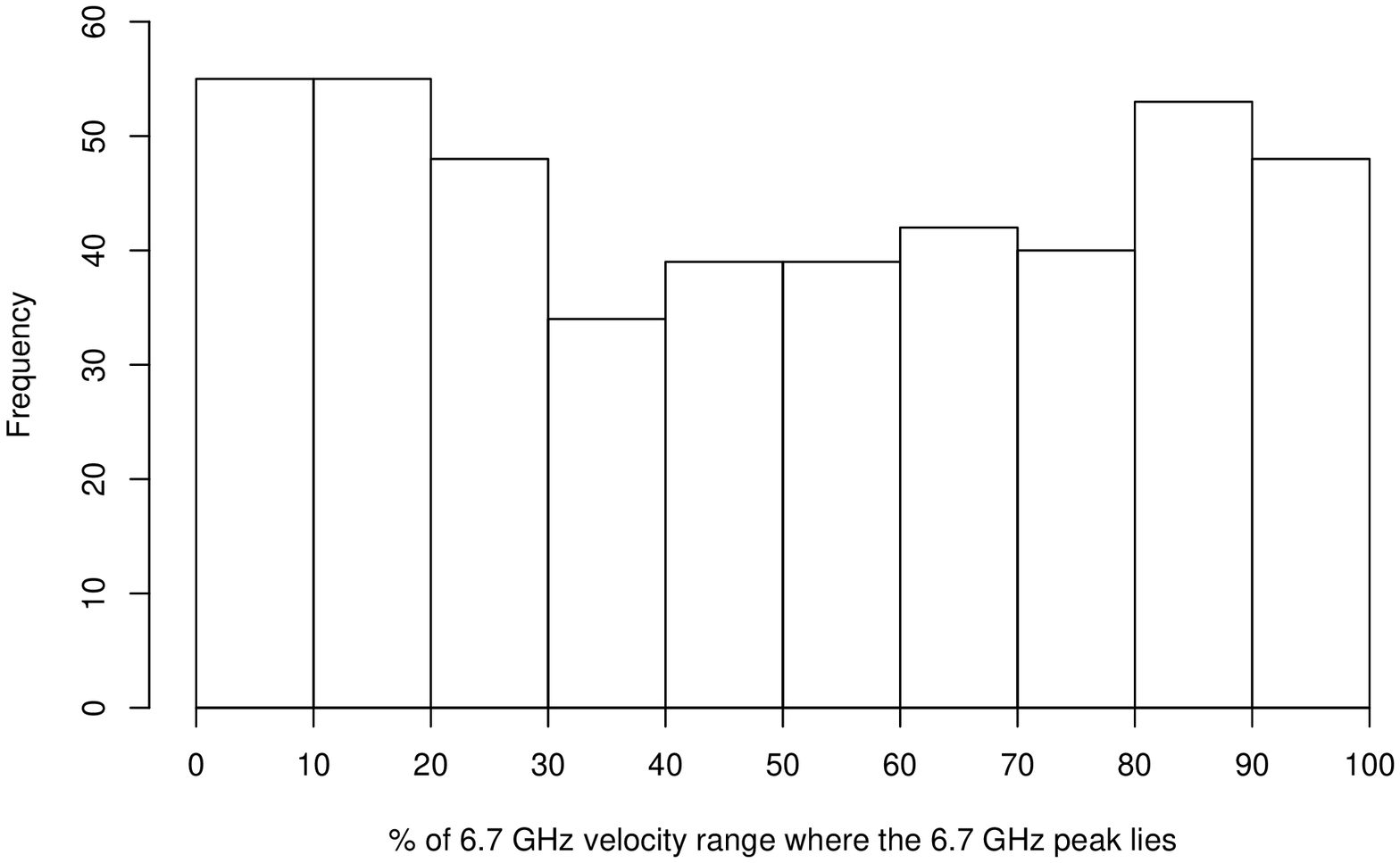}
\includegraphics[angle=270,scale=.40]{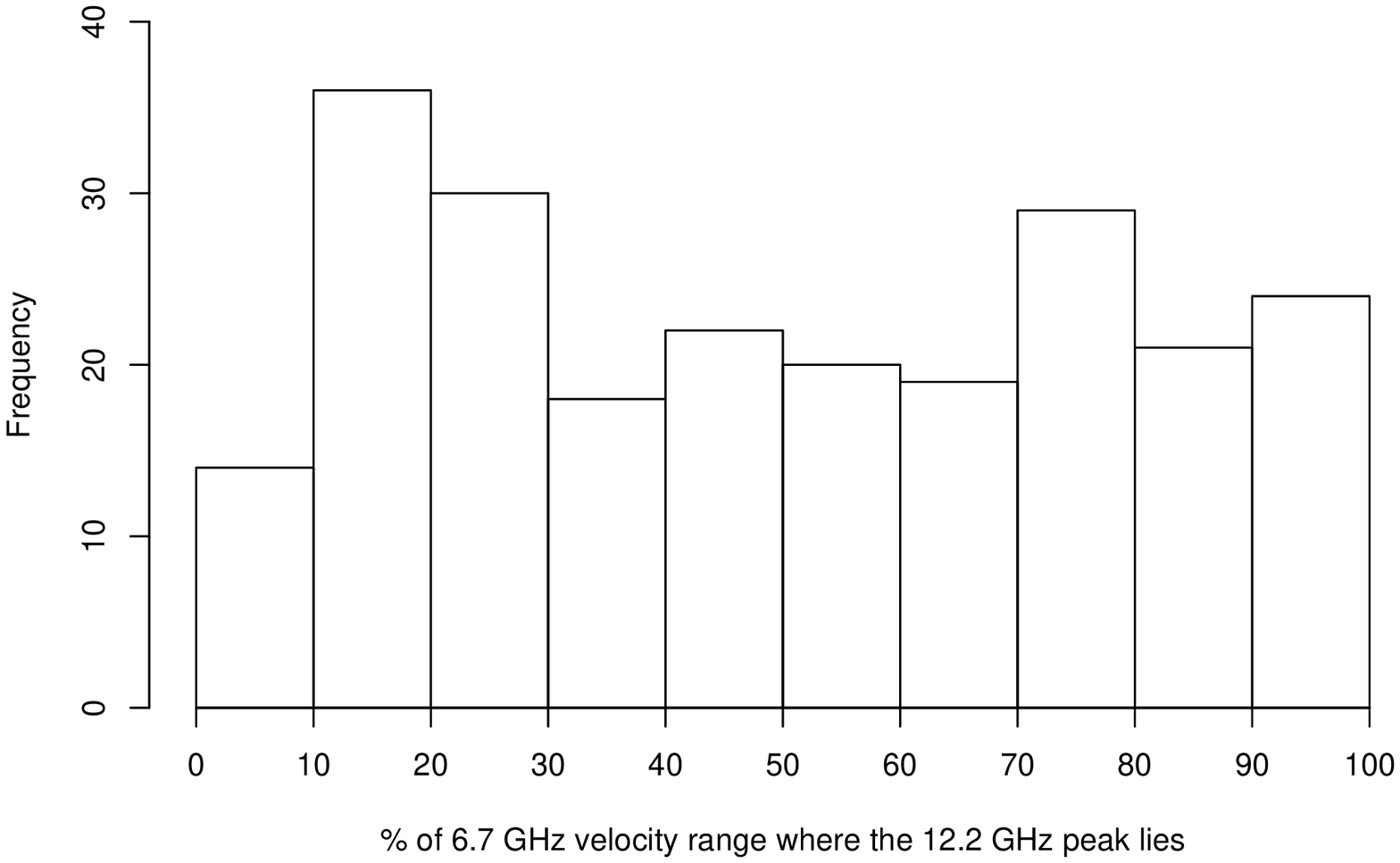}
\caption{Percentage of the 6.7~GHz methanol maser velocity range where the 6.7~GHz peak lies (top), and the 12.2~GHz lies (bottom). Only sources with 6.7~GHz methanol maser velocity ranges of more than 2 \kms are plotted. Here 0 \% corresponds to the minimum velocity and 100 \% corresponds to the maximum velocity.}
\label{fig:percent_vel_ranges}
\end{figure}

Fig.~\ref{fig:percent_vel_ranges} presents histograms that show the velocity of the 6.7 and 12.2~GHz methanol maser peak with respect to the velocity range of the 6.7~GHz methanol maser. Sources with 6.7~GHz methanol maser velocity ranges of less than 2~\kms have been excluded from this plot to ensure that the analysis is not dominated by single-feature sources. Percentages of the velocity ranges are calculated from the most negative velocity of the 6.7~GHz maser emission defined by the equation:

\begin{eqnarray*} 
\mbox{percentage}=100\frac{(\mbox{V$_{peak}$}-\mbox{V$_{min}$})}{(\mbox{V$_{max}$}-\mbox{V$_{min}$})}, 
\end{eqnarray*}

where V$_{min}$ and V$_{max}$ correspond to the minimum and maximum velocities of the 6.7~GHz methanol maser emission, and V$_{peak}$ corresponds to the velocity of the 6.7~GHz peak and the 12.2~GHz methanol maser peak in the top and bottom panels of Fig.~\ref{fig:percent_vel_ranges} repecitvely. These histograms show that there are slightly more sources whose peaks lie close to either edge of the velocity range than in the center, for both the 6.7 and 12.2~GHz sources.


We have compared the number of sources in each of the velocity range bins with the number of sources expected in each bin (taken to be even across the bins) and tested the significance with a chi-squared test. More 6.7~GHz methanol maser peaks lie close to the minimum velocity range, between 0 and 20~\%, than the expected number (43) which is almost statistically significant ({\em p}-value of 0.07). Furthermore, there are fewer 12.2~GHz methanol maser peaks that lie in the 0 to 10 \% bin of the 6.7~GHz methanol maser velocity range. The number of sources that fall in this bin (14) is almost statistically different from the number that would be expected (23). However, the number of 12.2~GHz methanol maser peaks that fall within 10 to 20~\% of the velocity range (measured from the minimum velocity) of the 6.7~GHz methanol maser is statistically significantly higher than the expected number ({\em p}-value 0.01). The combined bins between 0 and 20~\% show no significant deviation from the mean. 
Interestingly, the two bins in each histogram that either achieve statistical significance, or close to it, all fall close to the minimum velocity range. 

\clearpage

Although perhaps not formally significant, a slight excess of sources with peak emission blue-shifted with respect to the systemic velocity might result if there are some sources where blue-shifted emission is enhanced as a result of an outflow, and the presence of a central continuum source that is preferentially amplified by the emission on the side of the source moving towards us (as in maser models presented in \citet{Sob97}).

\subsection{Variability}\label{sect:var}

\citet{Caswell95b} showed that 12.2~GHz methanol masers commonly vary on time scales of three months by factors greater than 10~\%, but rarely at the extreme levels that are seen in some water masers. On comparing their observations with those of \citet{Cas93}, \citet{Caswell95b} noted that variations seen over several 3 month periods were similar to the variations over 4 years. This prompted \citet{Cas+95} to suggest that the sources may go through quasi-periodic fluctuations. Indeed, long term monitoring of some 12.2~GHz methanol masers have shown remarkable periodicity in intensity fluctuations \citep[e.g.][]{Goed05b,Goed09}. Observations of G\,9.62+0.02 lead \citet{Goed05a} to conclude, that at least in the case of this source, that the periodic flares were likely caused by periodic changes to the pump photon levels.

We carried out our 12.2~GHz observations over two epochs separated by 6 months, observing a large fraction of sources at both epochs. However, repeat observations of sources were preferentially targeted towards those 6.7~GHz sources where no 12.2~GHz methanol maser was observed in the first epoch, or those where only weak 12.2~GHz emission was detected. The incomplete nature of the sample of sources studied at both epochs prevents an extensive study of variability from being completed. Instead, sources with two epochs of data have been used to investigate the completeness of our search, as well as allowing a preliminary assessment of 12.2~GHz source variability.

While approximately half of the 6.7~GHz methanol maser targets were observed during both epochs (264 of 580), only 56 of these resulted in a detection during either epochs. 28 of these 56 sources could only be identified at one epoch, the strongest of which was 1.4~Jy when detected. The average flux density of these once-detected-sources was $\sim$0.6~Jy, which is comparable to the 5-$\sigma$ detection limit of our observations. However, as remarked in Section~\ref{sect:results}, many detections were confidently identified at the 3-$\sigma$ level. In addition to this, averaging the observations of the 28 sources over the two epochs resulted in a decreased signal-to-noise in the resultant spectra. This shows that these 28 weak sources detected at only one epoch, do exhibit real variations in peak flux density of close to 50~\% over the 6 month period: suggesting that the majority of these weak sources are more variable on average than the total 12.2~GHz methanol maser population.  The luminosities of the 6.7~GHz methanol masers (discussed in Section~\ref{sect:lum}) that are the counterparts of these 28 once-detected 12.2~GHz sources have an average luminosity that is almost five times lower than that of the full sample of 6.7~GHz masers with 12.2~GHz counterparts.


Given that half of the 56 sources that were observed during both epochs were detectable at only one epoch, it is likely that a significant number of additional 12.2~GHz methanol masers would be detected if a second epoch of observations were carried out for all sources. Of the 580 6.7~GHz methanol masers searched for 12.2~GHz methanol maser emission, 330 failed to yield detections and 122 of these were observed only once. Since 12 \% (28 of 236) of the sources that were observed at both epochs but only revealed emission on one occasion, we can estimate that $\sim$15 additional sources (12 \% of 122) would be detected if a second epoch of observations carried out towards the 122 sources that were only observed once and revealed no emission.


We fail to detect three 12.2~GHz methanol masers that were previously detected by \citet{Caswell95b} (G\,284.352--0.419, G\,327.291--0.578 and G\,349.067--0.017). These sources were reported to have peak flux densities of less than 1~Jy at the time of their detection. This is consistent with our estimation of the number of sources that would fall below our detection limits at any one epoch.


Of the 28 sources that were observed and detected at both epochs, the average variation in peak flux density was 20~\%. Only 8 of these sources had flux densities greater than 1~Jy at either epoch. Two sources showed variations of considerably more than 20~\%, at levels of 71 and 59~\% respectively. Ten of the 28 sources showed no variation over the two epochs. 

\subsection{Association with OH masers}\label{sect:OH}

The locations of the target 6.7~GHz methanol masers (with and without 12.2~GHz counterparts) have been compared with the locations of OH masers detected by \citet{C98}. \citet{C98} conducted a complete search of a portion of the Galactic plane,  along with a range of targeted observations for OH masers. This has allowed us to not only identify methanol masers with associated OH masers but also, for a large number of sources, those without. 393 of our 6.7~GHz maser target sample fall within the regions searched for OH masers by \citet{C98} and 144 of these are located within 1.5 arcsec of an OH maser. Using 1.5 arcsec as an association threshold takes into account the positional uncertainties for both sets of observations and therefore allows the majority of methanol-OH associations to be identified while minimizing the number of chance associations. Furthermore, given that OH maser spots excited by a single object
are often spread over regions greater than one arcsec \citep{FC89}, setting a tighter association threshold would not be appropriate. 

Both Lewis (2008) and \citet{Breen10a} found that 12.2~GHz methanol masers were preferentially detected towards 6.7~GHz methanol masers that are associated with OH masers. Given that these searches were potentially biased towards the more evolved 6.7~GHz methanol maser sources \citep{Breen10a}, we have tested this theory by comparing our complete methanol maser sample with the complete OH maser sample of \citet{C98}. 

We find, from our 12.2~GHz search towards our unbiased sample of 6.7~GHz methanol masers, that of the 393 6.7~GHz methanol masers with available OH data, 144 have associated OH masers and that 86 of these also have detections at the 12.2~GHz transition of methanol. Thus the search for 12.2~GHz towards 6.7~GHz methanol masers with OH masers has resulted in a detection rate of 60~\%. For the 249 6.7~GHz methanol maser sources with no associated OH maser, 99 have associated 12.2~GHz emission and therefore a search toward these 6.7~GHz sources has yielded a detection rate of 40~\%. An alternative statistic is that 46.5~\% of the 185 6.7~GHz methanol masers with 12.2~GHz counterparts (that fall within the regions completely searched for OH maser emission) have an OH maser counterpart. In comparison, the rate of OH maser association with the 28 weak 12.2~GHz methanol masers detected at only one epoch (mentioned in Section~\ref{sect:var}) is much lower with only one association out of 12 sources that fall within the region completely searched by \citet{C98}. This finding is consistent with these weak variable sources representing perhaps the youngest 12.2~GHz sources. 

In the regions covered by complete searches for both 6.7~GHz methanol masers and OH masers \citep{C98}, the number of methanol masers detections out-number OH detections by more than two to one. Many estimates suggest that the association rate for methanol masers towards OH masers may surpass 80~\% \citep{C09}. However it is clear from comparison of the regions covered by both the OH maser survey of \citet{C98} and the MMB catalog that the association rate for OH masers towards methanol masers is much lower, and in fact, is probably less than 40~\%. These number statistics may imply that the lifetime of methanol masers is much greater than that of the OH maser sources, but also shows that while the detection rate of 12.2~GHz methanol masers is greater at sites of OH masers, a search exclusively towards these sites would only uncover a fraction of the total population, since so many methanol masers are devoid of OH counterparts. The elevated 12.2~GHz detection rate towards sources associated with OH masers also lends support to the idea that the 12.2~GHz sources are present towards the  later stages of the period of evolution where 6.7~GHz masers are seen.

\subsection{Source luminosities}\label{sect:lumin}

Comparison between the luminosities of the two methanol maser transitions, and the other characteristics of each of the regions, has the potential to reveal much more insightful results than previous comparisons of only flux densities. \citet{Breen10a} identified a relationship between increasing 6.7~GHz methanol maser peak luminosity and evolution, whereby the more luminous 6.7~GHz methanol masers are associated with a generally later evolutionary stage of high-mass star formation. Comparison between ammonia linewidths and methanol maser luminosities by \citet{Wu10} show that the lower luminosity methanol masers correspond to the younger sources which is in agreement with the findings of \citet{Breen10a}. The analysis carried out by \citet{Breen10a} was based on the peak flux densities of the 6.7~GHz sources extracted from the literature rather than the more appropriate integrated flux density, which is expected to be a more robust estimate of a larger volume of space. Furthermore, in many cases the observations of the two methanol maser transitions considered in \citet{Breen10a} were separated by 10+ years. Since we have integrated flux densities for both of the methanol maser transitions available for our large sample of sources, here we repeat some of the analysis presented in \citet{Breen10a} and offer new insights from additional investigations.  

\subsubsection{Distances}

The calculated luminosity of the methanol masers are highly dependent on the adopted distance measurements. We have calculated the kinematic distances using the 6.7~GHz methanol maser peak velocities for our sample, using both the revised prescription offered by \citet{Reid09} (which uses a flat rotation curve and updated solar parameters), as well as that of more traditional model of \citet{BB93}. The agreement between the peak methanol maser velocity and molecular line data is within a few \kms for most 6.7~GHz sources and are therefore adequate for the computation of kinematic distances \citep{Szy07,Pandian09}.

Comparison of the distances calculated using the two different methods shows that they are in general very similar, although those calculated using \citet{Reid09} tend to be slightly smaller. As such, the choice of model does not affect the calculated luminosities greatly. There are a few instances where the derived kinematic distances are not plausible, and in these cases neither the model of \citet{BB93} or the prescription of \citet{Reid09} are able to give an acceptable distance, indicating that either these sources have anomalous velocities or they lie in poorly constrained regions of the Galaxy (especially when R $<$ 3.5 kpc, where extreme departures from circular rotation are known to be present \citep{CasMMB10}).


We adopt the prescription of  
\citet{Reid09} which achieves better correspondence with parallax  
measurements. Unfortunately sources with kinematic distance  
ambiguities have a more significant impact on the luminosity  
distribution. The accurate assignment to the near or far distance  
requires techniques such as parallax measurements or H{\sc i} self  
absorption, which are beyond the scope of this paper. For $\sim$15 \% of sources the calculated distance is unambiguous, for those with near/far ambiguities we have used the near distances in all analysis. Methanol maser sources with calculated distances of less than 0.5 kpc or more than 20 kpc have been removed from this analysis as they are likely to be erroneous. Where we believe that out assumption of the near distance have affected our results we mention it in the text.


\subsubsection{Comparison between 6.7~GHz luminosities and H$_{2}$ number densities}\label{sect:lum}

\begin{figure}
\includegraphics[angle=270,scale=.65]{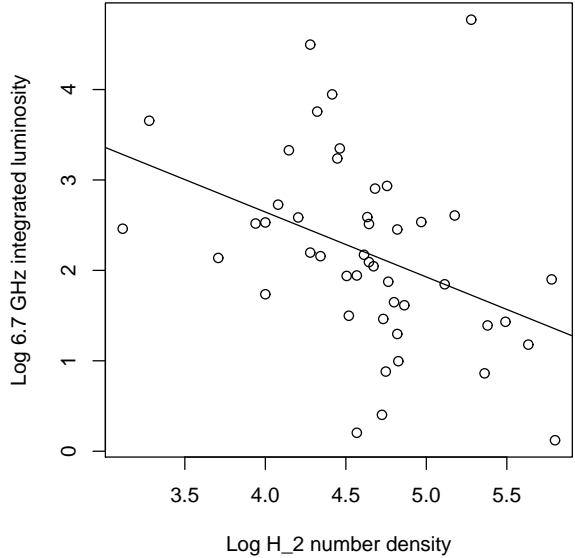}
\caption{Log of the 6.7~GHz integrated luminosity (Jy \kms kpc$^2$) versus the log of the H$_{2}$ number density of the associated 1.2 mm dust clump (cm$^{-3}$). The line of best fit is also plotted.}
\label{fig:density_6lum}
\end{figure}

\citet{Breen10a} found that there was a relationship between the H$_{2}$ number densities of 1.2~mm dust clumps (Hill et al. 2005) and the presence of \UCHII regions, with the more evolved 1.2~mm dust clumps (i.e. those with associated \UCHII regions) showing lower densities than the younger dust clumps. Plots of 6.7~GHz peak luminosity versus H$_{2}$ number density (fig.\ 2 of \citet{Breen10a})  showed that the more luminous 6.7~GHz methanol masers were associated with the less dense 1.2 mm dust clumps, implying a relationship between 6.7~GHz methanol source luminosity and evolution. Furthermore, as noted in \citet{Breen10a}, the appearance of the more evolved sources as being less dense may have been a consequence of the constant temperature assumed by \citet{Hill05}, and may be, at least in part, indicating that the more evolved sources are warmer. In their analysis, \citet{Breen10a} calculated the 6.7~GHz methanol maser luminosity using the peak flux density of the sources since the integrated flux densities were not available in the literature. Here we use the luminosities calculated from the integrated flux density to test whether the simpler analysis using luminosities calculated from the peak flux densities led to any bias.  Fig.~\ref{fig:density_6lum} shows a similar plot to fig.\ 2 of \citet{Breen10a} but uses the integrated flux densities of 6.7~GHz sources south of declination --20$^{\circ}$ (from the MMB observations) that overlap with the \citet{Hill05} 1.2~mm dust clump sample. Since the H$_{2}$ number densities were calculated using kinematic distances for each source as presented in Hill et al. (2005), (we have used the near distances in the case of ambiguities) to calculate the integrated 6.7~GHz luminosities for this plot for consistency.

Linear regression analysis was carried out on the data contained in Fig.~\ref{fig:density_6lum} and showed that the negative slope of the line of best fit of the distribution is statistically significant, with a weak correlation between individual data points (correlation coefficient of 0.39). The null hypothesis that the line of best fit has a slope of zero can safely be rejected ({\em p}-value 0.007). The line of best fit has the equation

\begin{eqnarray*}
  \mbox{log(L$_{6.7}$)}=-0.72[0.25](\mbox{log}(\mbox{H}_2))+5.52[1.17],
  \end{eqnarray*}
  
 where `L$_{6.7}$' is the integrated luminosity of the 6.7~GHz methanol maser (Jy~\kms pc$^2$) and `H$_{2}$' is the H$_{2}$ number density (cm$^{-3}$) calculated from the associated 1.2~mm dust clump emission. The numbers contained in the square brackets are the standard errors of the slope and the y-intercept respectively. As mentioned in \citet{Breen10a}, it is likely that at least some of the scatter present in the distribution of data points can be explained by the use of the near kinematic distances, which is an assumption prone to introducing errors. Additionally, since the 1.2~mm dust continuum observations of \citet{Hill05} used to determine the H$_2$ number densities are of relatively low resolution, we are comparing the maser positions with the broader properties of the associated region, as averaged by the large beam. A further contributor is likely to be the variable nature of the maser sources \citep[see e.g.][]{Goed}. The correlation between data points is slightly lower for sources shown in Fig.~\ref{fig:density_6lum} than those in fig. 2 of \citet{Breen10a}. However, given the smaller sample size in Fig.~\ref{fig:density_6lum} (46 compared to 113) it is difficult to draw meaningful conclusions about what this may mean. The much smaller sample size is a result of the fact that a large portion of the \citet{Hill05} sample (towards which the observations of \citet{Breen10a} were targeted) lie north of the declination cut-off imposed on the target sources. Regardless, this analysis shows that the result derived in \citet{Breen10a} holds when the 6.7~GHz integrated luminosities are used, confirming that, in general, 6.7~GHz methanol masers increase in luminosity as they evolve. 

\subsubsection{The relationship between the 6.7~and 12.2~GHz source luminosities}

\citet{Breen10a} identified a relationship between the ratio of 6.7 to 12.2~GHz peak flux densities with the luminosities of the respective data. The ratio of 6.7 to 12.2~GHz peak flux density apparently increased with increasing 6.7~GHz peak luminosity, implying that as the source evolved, the 6.7~GHz methanol maser increased in flux density more rapidly than the associated 12.2~GHz methanol maser. We have repeated that analysis on our much larger and complete sample using not only the peak flux densities of the sources, but also comparing the integrated flux densities. 

Log-log plots of: 

\begin{enumerate}
\item the ratio of 6.7~GHz to 12.2~GHz peak flux density ratio versus the 6.7~GHz integrated luminosity; 
\item the ratio of 6.7~GHz to 12.2~GHz methanol maser integrated flux densities versus the 6.7~GHz integrated luminosity; 
\item the ratio of 6.7~GHz to 12.2~GHz methanol maser peak flux densities versus the 12.2~GHz integrated luminosity, 
\end{enumerate}
were created and inspected. Together with linear regression analysis, these plots showed similar results to those reported in \citet{Breen10a}. In all cases the slopes of the line-of-best-fit were consistent with the notion that the luminosity of both of the methanol maser transitions increase with evolution and that the rate of increase is lower for the 12.2~GHz sources. 

Table~\ref{tab:lum_reg} shows a summary of the results from the linear regression analysis for the data corresponding to the three plots described above. From this table it can be seen that the correlation coefficients in all cases correspond to a weak correlation between individual data points and this scatter can largely be accounted for by the inaccuracies introduced by the assumption that all of the methanol masers lie at the near kinematic distance.

\begin{table}
\begin{center}
  \caption{Relationship between relative 6.7 and 12.2~GHz luminosities, searching for any dependence on 6.7 or 12.2~GHz integrated luminosities, and described in the text as plots 1, 2 and 3.}
  \begin{tabular}{ccccc}\hline\label{tab:lum_reg}
{\bf Plot} &         {\bf Slope}   &  {\bf Error	} &   {\bf {\em p}-value}   & {\bf R}\\  \hline
1.	& 0.11	& 0.03	&	2.28 $\times$ e$^{-4}$	& 0.23\\
2.	& 0.11	& 0.03	& 1.79 $\times$ e$^{-3}$	& 0.19\\
3.	& --0.12	&  0.03	& 6.24 $\times$ e$^{-5}$	& 0.25\\ 
\hline
	\end{tabular}
\end{center}
Column 2 shows the slope of the line of best fit, column 3 shows the standard error associated with the slope of the line, column 4 shows the {\em p}-value relating to the null hypothesis that the line has a zero slope, and column 5 gives the correlation coefficient.
\end{table}

\begin{figure}
\includegraphics[angle=270,scale=.65]{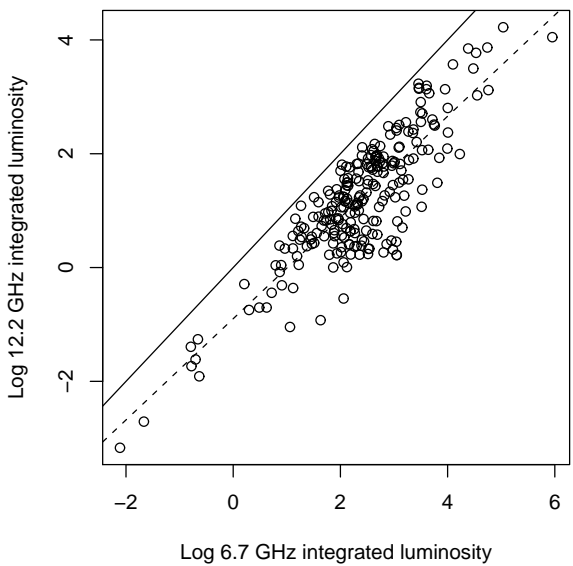}
\caption{Log 12.2~GHz integrated luminosity versus the log of the 6.7~GHz integrated luminosities (luminosities in units of Jy \kms kpc$^2$). Overlaid is the line of best fit (dashed) and line with a slope of 1 (solid).}
\label{fig:6lum_12lum}
\end{figure}

\begin{figure}
\includegraphics[angle=270,scale=.65]{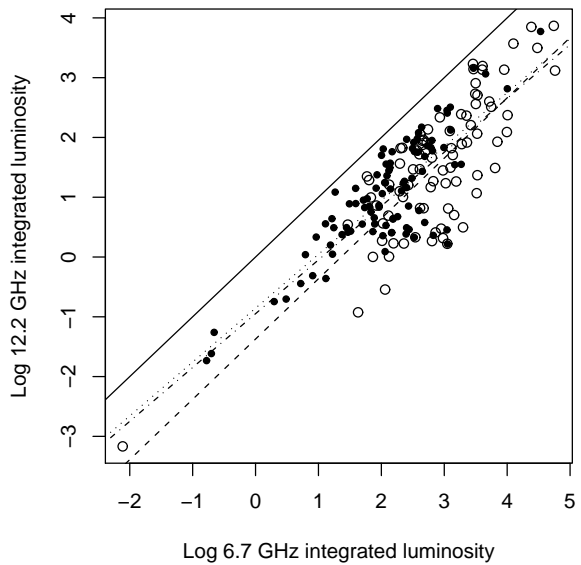}
\caption{Log 12.2~GHz integrated luminosity versus the log of the 6.7~GHz integrated luminosities (luminosities in units of Jy \kms kpc$^2$). The unfilled circles represent those sources with an associated OH maser and the filled circles show the sources with no OH maser counterpart. Overlaid is the line of best fit for sources with OH masers (dashed), sources without OH masers (dotted), the full sample with OH information (dot-dashed) and line with a slope of 1 (solid) for comparison. A summary of the linear regression analysis for each of the source categories is shown in Table~\ref{tab:lum_oh}.}
\label{fig:6lum_12lum_oh}
\end{figure}

A comparison between the 6.7~GHz and 12.2~GHz luminosities calculated from their integrated flux densities is shown in Fig.~\ref{fig:6lum_12lum}. Not surprisingly, this figure shows that the majority of these methanol maser sources have intermediate luminosities, evident by the concentration of sources in the log luminosity range 1 to 3 and 0 to 2 for the 6.7~GHz and 12.2~GHz sources respectively. The relationship between the 6.7 and 12.2~GHz methanol maser luminosities is well estimated by the line of best fit as it shows a relatively tight correlation (correlation coefficient is 0.85) between individual data points, with no extreme outliers. This plot similarly shows that for higher 6.7~GHz luminosities, the ratio of 6.7~GHz to 12.2~GHz source flux density increases. This is consistent with the maser models presented in \citet{Cragg05} wherein fig. 2 shows that the 6.7:12.2~GHz flux density increases as the source evolves. The line of best fit has the equation

\begin{eqnarray*}
  \mbox{log(L$_{12.2}$)}=0.89[0.03](\mbox{log}(\mbox{L$_{6.7}$}))-0.90[0.09],
  \end{eqnarray*}
  
 where `L$_{12.2}$' is the integrated luminosity of the 12.2~GHz methanol maser  and `L$_{6.7}$' is the integrated luminosity of the 6.7~GHz methanol maser. The numbers contained in the square brackets are the standard errors of the slope and the y intercept respectively. 

We have further investigated this relationship by splitting the data into two categories. The first category includes all 6.7 and 12.2~GHz methanol masers with associated OH masers \citep{C98} and the second includes those sites with both 6.7 and 12.2~GHz methanol masers but no associated OH maser. Fig.~\ref{fig:6lum_12lum_oh} presents a plot similar to Fig.~\ref{fig:6lum_12lum} but with methanol sources with associated OH masers distinguished from those without. Methanol sources that have not been sensitively searched for OH masers have been excluded from the plot (a total of 58 sources). Interestingly, the methanol sources with OH maser counterparts tend towards the higher luminosities, lending further support to the idea that methanol masers tend to increase in luminosity as they evolve (since OH masers are known to be associated with a later stage of evolution \citep{FC89,Cas97}).

The summary of the linear regression analysis carried out on the three categories of sources displayed in Fig.~\ref{fig:6lum_12lum_oh} (all sources, sources with OH and sources with no OH counterpart) are presented in Table~\ref{tab:lum_oh}. The characteristics computed in the linear regression analysis for all sources shown in Fig.~\ref{fig:6lum_12lum_oh} are almost identical to those derived for the complete sample (Fig.~\ref{fig:6lum_12lum}). Sources with associated OH masers seem to have a slightly more positive slope ($\sim$1) than the sources with no OH maser counterparts that have a slope equal to that of the full sample (0.90). Taking into account the standard errors associated with these two lines shows that they could have the same slope (i.e. both less than one), although the possibility that this is a real difference is intriguing. If the difference is real, it implies that although the methanol maser luminosity increases more rapidly in the 6.7~GHz tranisition as the source evolves, towards the evolutionary stage where OH masers are present this effect lessens.

\begin{table}
\begin{center}
  \caption{Summary of results from linear regression analysis for data presented in Fig.~\ref{fig:6lum_12lum_oh}. }
  \begin{tabular}{ccccc}\hline\label{tab:lum_oh}
{\bf Category} &         {\bf Slope}   &  {\bf Error} &   {\bf {\em p}-value}   & {\bf R}\\  \hline
full sample	& 0.90	& 0.05	& $<$2 $\times$ e$^{-16}$	& 0.82\\
with OH		& 1.01	&	0.08	& $<$2 $\times$ e$^{-16}$	& 0.80\\
no OH		& 0.90	& 0.06	& $<$2 $\times$ e$^{-16}$	& 0.83\\
	\end{tabular}
\end{center}
Column 1 gives the category of the sources in the sample, column 2 shows the slope of the line of best fit, column 3 shows the standard error associated with the slope of the line, column 4 shows the {\em p}-value relating to the null hypothesis that the line has a zero slope, and column 5 gives the correlation coefficient.
\end{table}

\subsubsection{Preliminary analysis of the luminosity of individual maser features}\label{sect:individualf}

A preliminary analysis has been carried out on individual features of 6.7~GHz masers with respect to their 12.2~GHz counterparts. Since the two methanol maser transitions are typically co-spatial to within a few miliarcseconds \citep{Mos02}, and that this is especially likely when the spectral profiles of the respective transitions are similar \citep{Norris93}, we are able to meaningfully draw these comparisons even in the absence of detailed high-resolution maps. Future studies of these sources will include an analysis of the individual features of a much large group of sources. 

Analysis of the individual features associated with each maser source has the potential to reveal much more about the sources, as well as the individual features, which show emission at both methanol transitions, or are devoid of a 12.2~GHz counterpart. Since the age of the exciting star will be common to individual maser features, their analysis will reap greater information on the study of the relative evolutionary stage that each methanol transition is associated with. Furthermore, insights into the conditions favourable to both transitions can be achieved by carrying out the analysis on such a large sample of sources.

\begin{figure}
\includegraphics[angle=270,scale=.65]{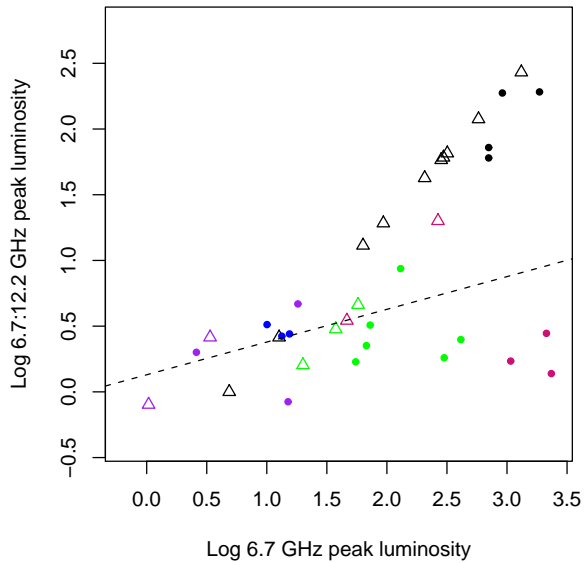}
\caption{Log of the 6.7:12.2~GHz individual feature ratio versus the log 6.7~GHz peak luminosity for every spectral feature associated with five methanol maser sources. Dots represent 6.7~GHz features with a 12.2~GHz counterpart, while triangles show features with no detectable 12.2~GHz counterpart, thus showing the lower limit of the 6.7:12.2~GHz feature ratio. Colours distinguish between individual maser sites. The dashed line shows the line of best fit from the sample of sources presented in \citet{Breen10a}, fig.~2, where a plot of 6.7~GHz peak luminosity versus 6.7:12.2~GHz peak ratio is shown for the peak 6.7~GHz feature associated with each source.}
\label{fig:6lum_ratio_features}
\end{figure}

Fig.~\ref{fig:6lum_ratio_features} displays the peak luminosity of each feature associated with five 6.7~GHz methanol masers, versus the ratio of 6.7 to 12.2~GHz luminosity of the corresponding feature. The five sources presented in this plot all exhibit three or more 12.2~GHz spectral features and were chosen randomly from this population. In addition to the ratios calculated from detections at both transitions (in individual features), upper limits on the ratios of 6.7~GHz features with no 12.2~GHz counterpart, are shown. The individual features of the five sources are colour-coded, revealing several trends.

Firstly, the 6.7~GHz luminosity versus the ratio of 6.7 to 12.2~GHz luminosity of each spectral feature are clustered together in groups containing all the features associated with the same source. It is expected from Section~\ref{sect:lum} (in conjunction with results presented in \citet{Breen10a}) that the ratio of 6.7 to 12.2~GHz luminosity will increase with 6.7~GHz luminosity, and therefore evolution of the source. The clustering of the features associated with each source supports this idea since the exciting source will be the same, and therefore at the same evolutionary stage, for all spectral features.

A second implication from Fig.~\ref{fig:6lum_ratio_features} comes from the location of the lower limits for the ratios. It can be seen that, while the features which show emission at both 6.7 and 12.2~GHz are clustered, the upper limits for features with no detectable 12.2~GHz counterpart lie on the outskirts of the cluster; since these are upper limits, if anything, they are likely to lie even further from the cluster. This implies that many of the features with no 12.2~GHz counterpart that are present on this plot are unlikely to be just below the detection limits of the search.

Also shown in Fig.~\ref{fig:6lum_ratio_features} is the line of best fit calculated from the linear regression carried out on the larger group of sources shown on a similar plot in \citet{Breen10a}, fig.~2 which shows only the peak feature for each of the methanol maser sources. The plotted line is the same (to within the errors) as the line of best fit calculated using the points plotted on Fig.~\ref{fig:6lum_ratio_features} (omitting the upper limits).

\subsubsection{Methanol maser luminosity and velocity range comparison}


There is a striking difference in the values of both luminosity and velocity range of the methanol masers in the association categories: firstly, the 6.7~GHz sources devoid of both 12.2 and OH maser counterparts; secondly, those with associated 12.2~GHz emission but not OH; and thirdly those with both 12.2~GHz and OH maser counterparts. We find that the 12.2~GHz methanol maser sources with associated OH masers have higher luminosities and velocity ranges than those with no OH counterpart. OH maser associations (or lack thereof) with the methanol masers have been determined from comparisons with the OH maser sources listed in \citet{C98}.

Table~\ref{tab:lum_vrange} shows the average and median velocity ranges and luminosities of both the 6.7 and 12.2~GHz methanol masers broken up in categories according to their associations. In the top half of the table the characteristics of the 6.7~GHz methanol masers are given and show a progressive widening of velocity range and increasing luminosity from the category which includes 6.7~GHz methanol masers with no 12.2~GHz methanol or OH maser counterpart; to 6.7~GHz methanol masers with associated OH maser emission but no 12.2~GHz methanol maser emission; to 6.7~GHz methanol masers with associated emission at the 12.2~GHz transition but no associated OH maser; and finally those 6.7~GHz methanol masers with both 12.2 and OH maser counterparts. The bottom half of the table gives the characteristics of the 12.2~GHz methanol masers with and without associated OH masers. In this case, a similar trend is seen, with the sources without OH masers showing smaller 12.2~GHz maser velocity ranges and lower luminosities than those 12.2~GHz sources with associated OH masers.

\begin{table*}
\begin{center}
  \caption{Average and median velocity ranges and luminosities of 6.7 and 12.2~GHz methanol masers broken into categories on the basis of their methanol and OH maser associations. }
  \begin{tabular}{lccccc}\hline\label{tab:lum_vrange}
 {\bf Associations}	& {\bf Sample} &	\multicolumn{2}{c}{\bf Velocity range} & \multicolumn{2}{c}{\bf Integrated luminosity }\\
& {\bf size} &	\multicolumn{2}{c}{\bf (\kmsns)} & \multicolumn{2}{c}{\bf (Jy \kms kpc$^{2}$)}\\
 && {\bf average} & {\bf median} & {\bf average} & {\bf median}  \\ \tableline
{\bf 6.7~GHz}\\
no 12.2 or OH 	&136&	5.2	&	3.9	&	70	&	25	\\
no 12.2 but OH	&56 &6.1 	&5.5	&  269 & 64\\
12.2 but no OH &93 & 8.3	& 7.6	& 	465	&	142	\\
12.2 and OH	&84 &	10.0	&	9.4	& 3819	&	892	\\ \hline
{\bf 12.2~GHz}\\
no OH		& 93 &	3.1	&1.6	& 35	&17\\
OH			& 84 &	4.7	&	3.0	&	667	&73\\ \hline
	\end{tabular}
\end{center}
In the top half of the table, characteristics of the 6.7~GHz methanol masers are shown, broken into their association categories, while the second half of the table presents the characteristics of the 12.2~GHz methanol masers. Column 1 specifies the association category for the 6.7 (top) and 12.2~GHz (bottom) methanol masers; column 2 gives the number of sources in each category; column 3 gives the average velocity range (\kmsns); column 4 gives the median velocity range (\kmsns); column 5 gives the average luminosity (Jy \kms kpc$^{2}$); and column 6 gives the median luminosity (Jy \kms kpc$^{2}$).
\end{table*}

\begin{figure*}\vspace{-1cm}
\includegraphics[angle=270,scale=.70]{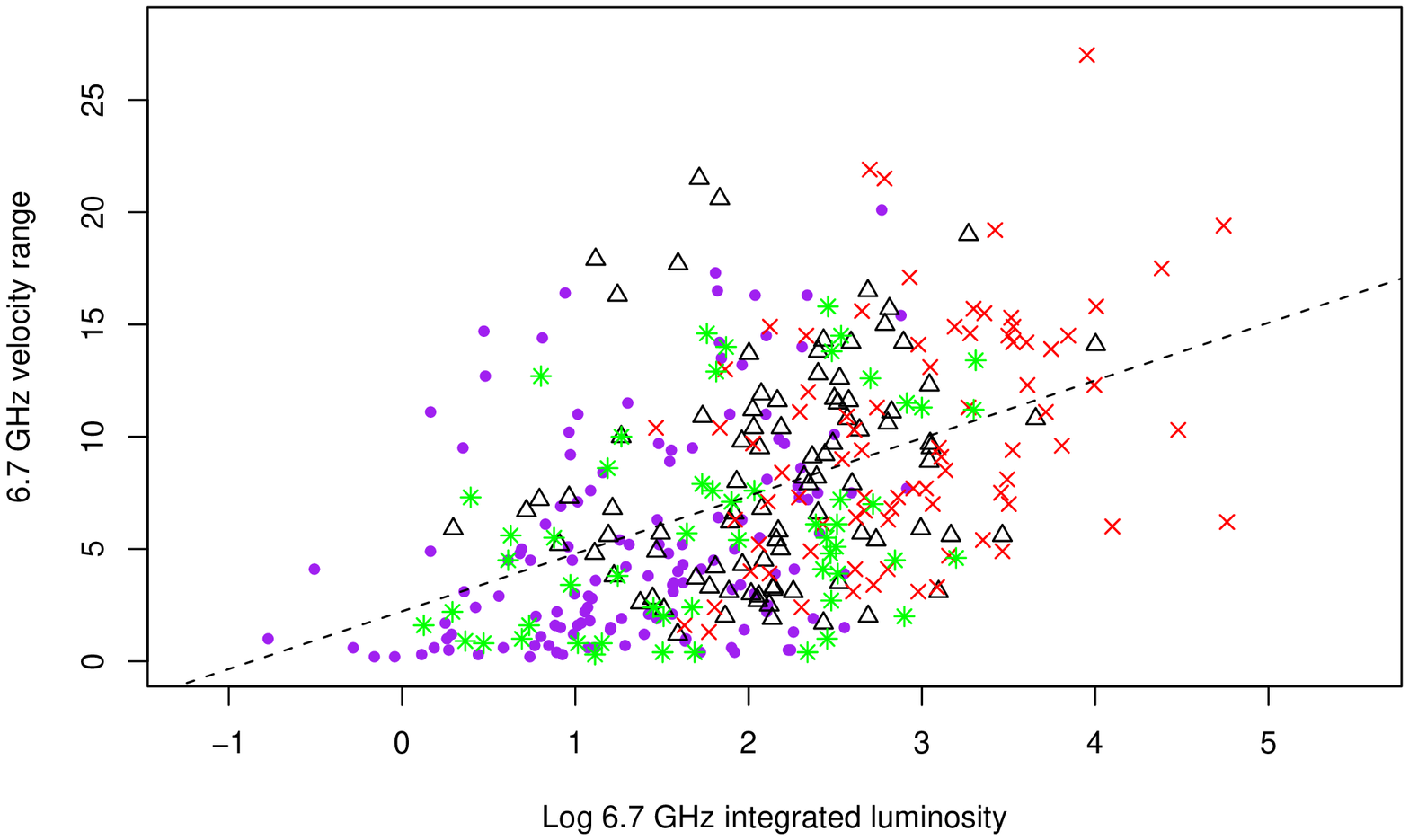}\vspace{-1.5cm}
\includegraphics[angle=270,scale=.70]{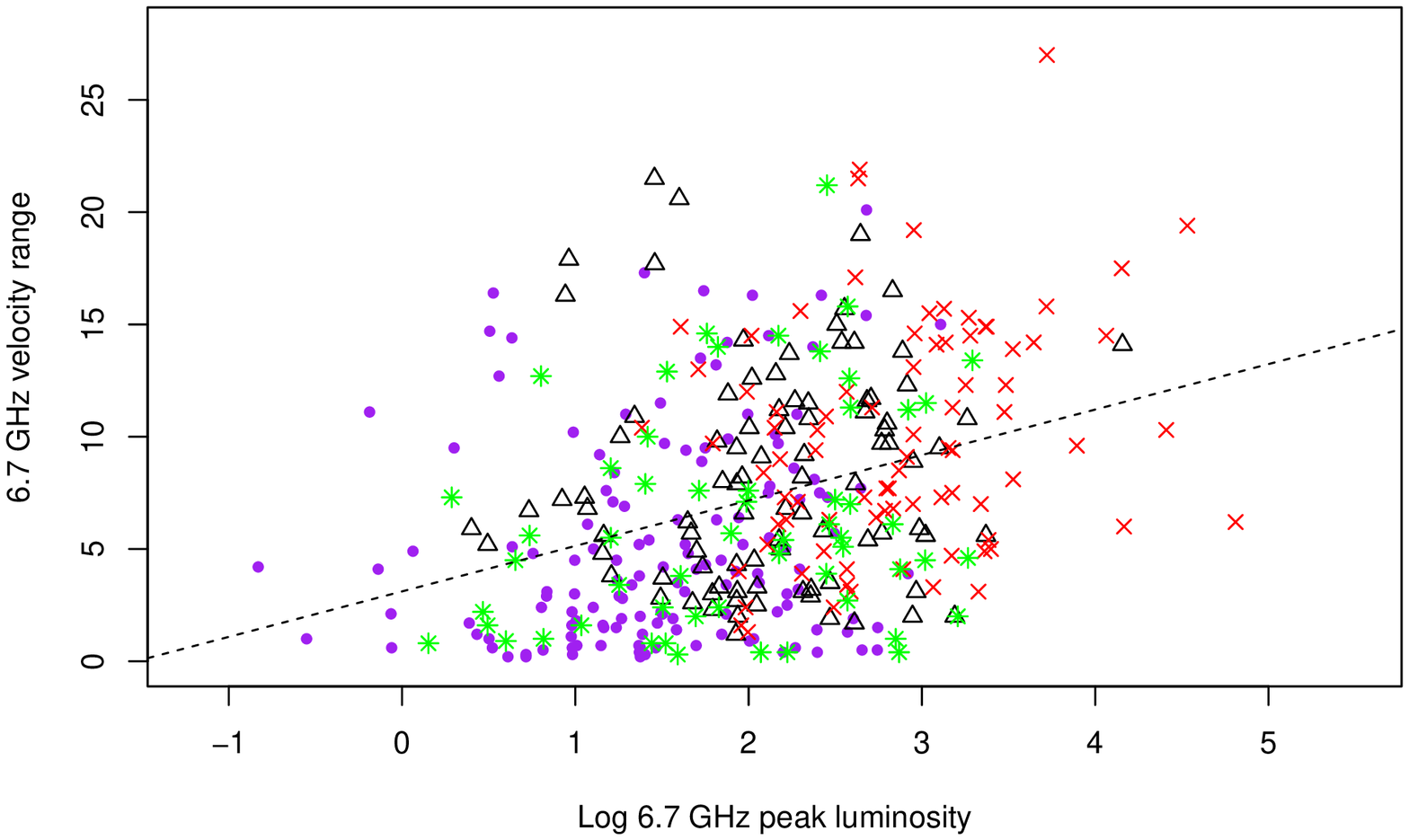}
\caption{Log of the 6.7~GHz integrated luminosity (Jy \kms kpc$^2$) versus the 6.7~GHz methanol maser velocity range (\kmsns) (top) and log of the 6.7~GHz peak luminosity (Jy kpc$^2$) versus the 6.7~GHz methanol maser velocity range (\kmsns) (bottom). To be present on these plot sources must have acceptably determined kinematic distances and integrated/peak flux densities. Shown on the plots are four categories of sources: 6.7~GHz sources with no associated 12.2~GHz methanol or OH masers (136 purple dots); 6.7~GHz methanol masers with detectable 12.2~GHz methanol masers but no associated OH masers (93 black triangles); 6.7~GHz methanol masers with both 12.2 methanol and OH maser counterparts (81 red crosses); and the 6.7~GHz methanol masers with no 12.2~GHz methanol maser counterpart but with OH maser emission (56 green asterisks). Overlaid are the lines of best fit that corresponds to all of the 6.7~GHz masers presented on each plot (linear regression gives: (top plot) slope=2.57[0.27], {\em p}-value below 2 $\times$ e$^{-16}$, and a correlation coefficient of 0.48; (bottom) slope=2.03[0.29], {\em p}-value=6.31 $\times$ e$^{-12}$, and a correlation coefficient of 0.34).}
\label{fig:6lum_vrange}
\end{figure*}

\begin{figure*}\vspace{-1cm}
\includegraphics[angle=270,scale=.70]{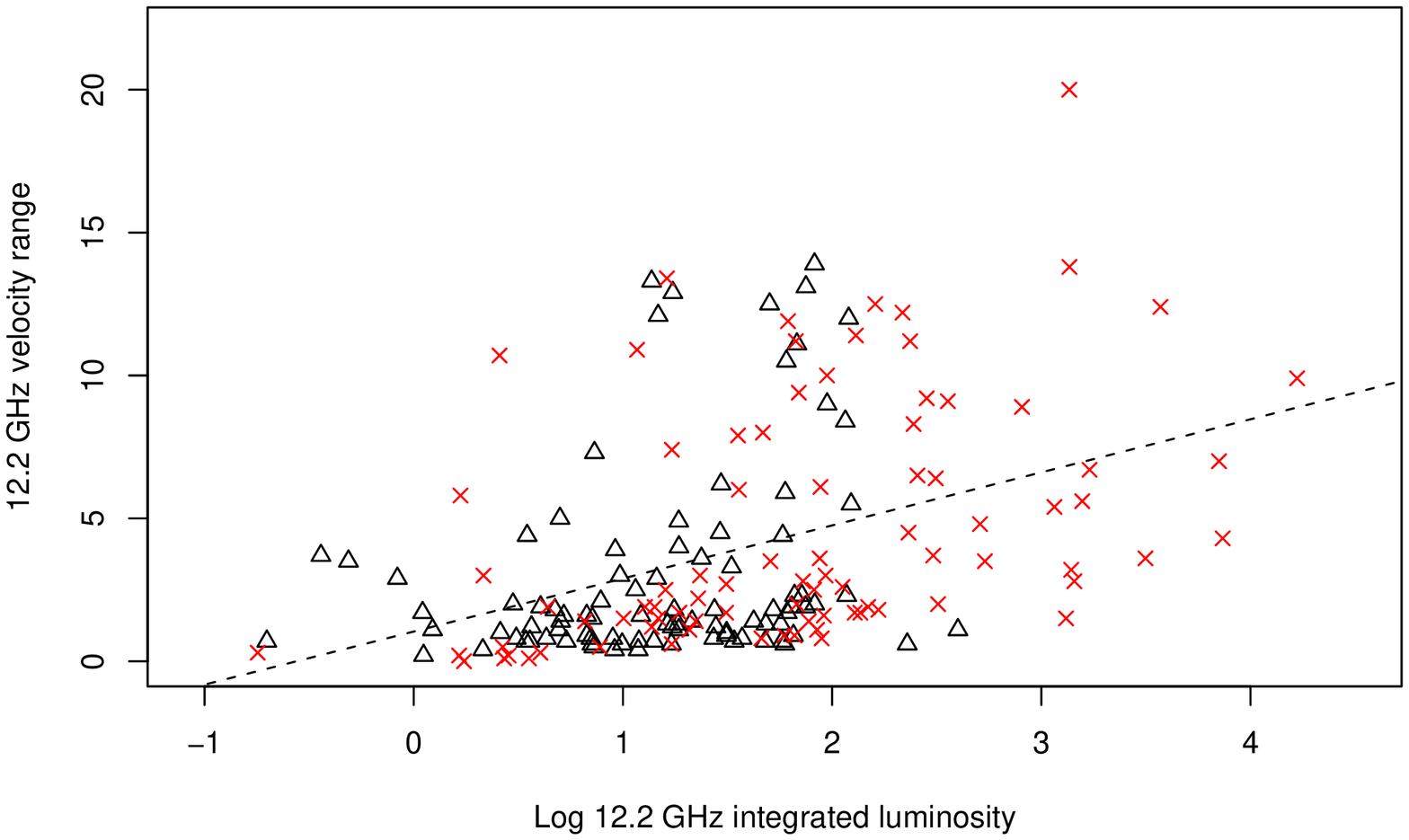}\vspace{-1.5cm}
\includegraphics[angle=270,scale=.70]{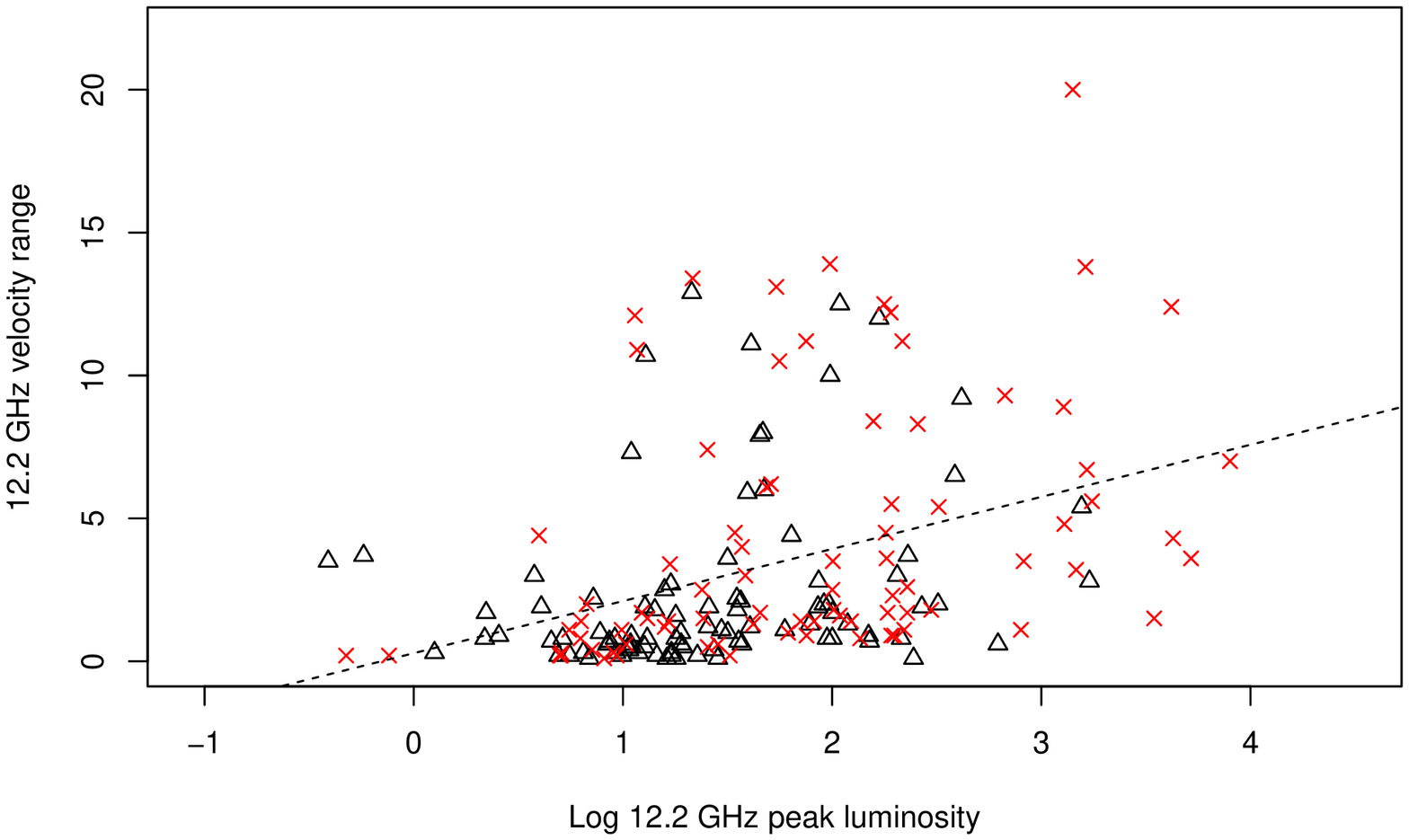}
\caption{Log of the 12.2~GHz integrated luminosity (Jy \kms kpc$^2$) versus the 12.2~GHz methanol maser velocity range (\kmsns) (top) and log of the 12.2~GHz peak luminosity (Jy kpc$^2$) versus the 12.2~GHz methanol maser velocity range (\kmsns) (bottom). Shown on the plots are 12.2~GHz sources with no associated OH masers (93 black triangles); and 12.2 methanol masers with OH maser counterparts (84 red crosses). Overlaid are the lines of best fit that corresponds to all of the 12.2~GHz masers presented on each plot (linear regression gives: (top) slope=1.86[0.31], {\em p}-value=1.72 $\times$ e$^{-8}$, and a correlation coefficient of 0.40; (bottom) slope=1.82[0.33], {\em p}-value=1.40 $\times$ e$^{-7}$, and a correlation coefficient of 0.38).}
\label{fig:12lum_vrange}
\end{figure*}

Figs~\ref{fig:6lum_vrange} and~\ref{fig:12lum_vrange} show plots of log 6.7~GHz luminosity versus 6.7~GHz velocity range and log 12.2~GHz luminosity versus 12.2~GHz velocity range, respectively. In these figures, there is one plot of the maser luminosity calculated using the integrated flux density and one using the maser peak flux densities. Fig.~\ref{fig:6lum_vrange} shows that there is an overall trend between increasing 6.7~GHz methanol maser luminosity (for both integrated and peak) and velocity range, with a significantly positive slope and a moderate correlation between data points. These plot also highlight the segregation of the values corresponding to the different association categories. Like \citet{Breen10a}, we argue that, in general, more evolved 6.7~GHz methanol masers exhibit higher luminosities. The values reported in Table~\ref{tab:lum_vrange} along with the data shown in Fig.~\ref{fig:6lum_vrange} support this idea. This plot offers additional insight, with a clear tendency for the more evolved 6.7~GHz methanol masers to show larger velocity ranges. The figure showing the luminosities calculated from the integrated flux densities suggests that the gas volume conducive to 6.7~GHz methanol maser emission also increases as the source evolves. Alternatively, the increase in velocity range may, at least partially, be attributed to increased internal motions within the more evolved sources.

There is, of course, the possibility that this relationship is influenced by sensitivity, i.e. for example, that the 6.7~GHz sources with associated 12.2~GHz methanol and OH masers have higher flux densities, increasing the likelihood that a more extensive velocity range would be identified. However, Fig.~\ref{fig:6lum_ratio_features} suggests that sensitivity biases are not the culprit for the difference between the sources. Even if this is the case for some sources, Fig.~\ref{fig:6lum_ratio_features} shows that the characteristics of the 6.7~GHz methanol masers change with evolution regardless (since this shows that the flux density increases). Furthermore, if sensitivity limitations were precluding us from observing the full extent of the 6.7~GHz methanol maser velocity range, then we would expect this to affect all of the association categories equally since the different categories will be interspersed through a range of distances in the Galaxy. A sample of the outlying sources shown in top plot of Fig.~\ref{fig:6lum_vrange} have been investigated, specifically those sources represented by purple dots and black triangles centred on $\sim$1 on the x-axis and  $\sim$17 on the y-axis. These sources show unexpectedly large velocity ranges for their calculated luminosities. An investigation of these sources shows that they all have low flux densities ($<$4~Jy) and this together with their location on the plot implies that they might actually lie at the far distance rather than the near distance that has been assumed. Therefore, a comparison between the integrated flux densities and velocity ranges of sources may provide a viable, indirect way of resolving near/far distance ambiguities.

A further group of sources are presented in Fig.~\ref{fig:6lum_vrange} and Table~\ref{tab:lum_vrange}; these are those 6.7~GHz methanol masers with associated OH maser emission but no 12.2~GHz methanol maser emission (represented by green asterisks). These sources do not fit in with the evolutionary sequence that seems to describe the majority of our sources. The 6.7~GHz methanol sources in this category have average and median velocity ranges and luminosities that fall in between the categories of `no 12.2 or OH' and `12.2 but no OH'. It seems, therefore, that for some sources, the OH masers switch on prior to the 12.2~GHz methanol masers. It can be seen that these sources basically populate the same range as that of the two categories of `no 12.2 or OH' and `12.2 but no OH'. These sources need further investigation to understand their nature. Of particular interest would be to compare the abundances of methanol in these sources compared to the other categories, as well as determining their associations with \UCHII and hyper-compact \ionhy regions.

Fig.~\ref{fig:12lum_vrange} displays the characteristics of the 12.2~GHz methanol masers, broken into the categories of sources associated with an OH maser (red crosses) and sources with no OH maser counterpart (black triangles). These plots similarly show evidence that the 12.2~GHz methanol masers increase in velocity range and luminosity as they evolve, since OH masers are found at a later stage of high-mass star formation \citep{FC89,Cas97}. The congregation of a number of sources with small velocity ranges and the log of the integrated luminosity between 0 and 2, perhaps shows that the 12.2~GHz sources are more affected by sensitivity limitations prohibiting the full velocity ranges to be identified. The sensitivity limits of the 6.7 and 12.2~GHz searches were similar, and together with the expectation that the associated 12.2~GHz methanol masers sources are weaker, means that this is not unexpected. However, analysis of individual features associated with five 6.7 and 12.2~GHz methanol masers shown in Section~\ref{sect:individualf} implies that, at least for these sources, the 6.7~GHz  features with no detectable 12.2~GHz counterpart are unlikely to be a result of sensitivity limitations. Both scenarios are probably true for some number of sources.

From analysis of the ratio of integrated flux densities of the two methanol transitions in Section~\ref{sect:comp}, it was suggested that there was a larger volume of gas that was conducive to 6.7~GHz methanol than their 12.2~GHz counterparts. Data presented in Table~\ref{tab:lum_vrange} and Figs~\ref{fig:6lum_vrange} and~\ref{fig:12lum_vrange} supports this, with a clear tendency for the 12.2~GHz methanol maser sources to have less extensive velocity ranges. In addition, this table and figures show strong evidence for the gas volume to increase as the sources evolve (for both the 6.7 and 12.2~GHz sources). In Fig.~\ref{fig:6lum_12lum}, and discussion thereof, we show that the luminosities of both the 6.7~GHz methanol maser and the 12.2~GHz methanol masers increased with evolution, but the the rate of increase was much greater for the 6.7~GHz sources. The slopes of the lines derived from linear regression analysis of the data presented in Figs~\ref{fig:6lum_vrange} and~\ref{fig:12lum_vrange} show that a similar effect is seen in the velocity ranges of the two methanol transitions. The slope associated with the 6.7~GHz methanol sources is 1.4 times that associated with the 12.2~GHz sources, showing that the velocity range of the 6.7~GHz methanol sources increases more rapidly than that of their 12.2~GHz methanol maser counterpart.

These findings support the idea that the different combinations of 6.7~GHz methanol, 12.2~GHz methanol and OH masers, trace differing evolutionary stages of high mass star formation in the following sequence (starting from youngest); 6.7~GHz methanol masers devoid of both 12.2~GHz counterparts and OH masers; 6.7~GHz methanol masers with 12.2~GHz counterparts but no associated OH maser emission; and finally, 6.7~GHz methanol masers with both 12.2~GHz counterparts and associated OH masers. This interpretation is self consistent with all other analyses completed and is also consistent with the independent results of \citet{Breen10a}. Further to this result, we find that not only the luminosity of the 6.7~GHz methanol maser increases more rapidly than that of the associated 12.2~GHz maser, but also the 6.7~GHz velocity range. This supports an evolutionary scenario whereby both the luminosity and the gas volume conducive to the 6.7 and 12.2~GHz masers increases as they evolve (although at a lesser rate for the 12.2~GHz sources).

\subsection{Summary of evolutionary results}

In this section we give a summary of the main evolutionary arguments presented in the previous sections and highlight the consistency of the these results. Our findings are consistent with those that are well established in the literature; particularly that OH masers are associated with a later stage of evolution \citep{FC89,Cas97}. Also, as previously mentioned, \citet{Wu10} found that low luminosity 6.7~GHz methanol masers are associated with small ammonia linewidths, which \citet{Longmore07} suggests are indicative of young source; consistent with the methanol masers increasing in flux density with evolution as suggested by \citet{Breen10a} and consistent with our current findings.

Below we list some of integral findings and their evolutionary interpretations, followed by some of the main supporting evidence for each of the statements.

{\bf 12.2~GHz detection rate is higher towards the more evolved sources.} This is evident from our comparison of detection rates towards sources with and without OH maser emission, but also in the high detection rates of previous searches which were biased towards more evolved star formation regions. This suggests that 12.2~GHz methanol masers are associated with second half of the 6.7~GHz methanol maser lifetime. This is further supported by the fact that there are few associations between OH masers and very weak 12.2~GHz methanol masers (which we interpret as being the youngest 12.2~GHz sources; see the next point).

{\bf The methanol maser emission increases in luminosity with evolution.} Comparison between the H$_2$ number densities calculated by \citet{Hill05} and methanol maser luminosity (both here and in \citet{Breen10a}) shows a trend that is consistent with the sources increasing in luminosity with evolution. Comparison between the luminosity of the two methanol maser transitions shows that the 6.7~GHz methanol maser emission is increasing at a higher rate than the 12.2~GHz sources. Furthermore, sources that show methanol maser have higher luminosities when they are also associated with an OH maser.

{\bf The methanol maser emission have wider velocities ranges as they evolve.} Luminosity of both methanol maser transitions increases with increasing velocity range, showing that the velocity range of sources is also increasing with evolution. Furthermore, both the velocity range and luminosity of sources increases in the association groups. In the case of the 6.7~GHz sources this increase is from solitary 6.7~GHz sources, to those with 12.2~GHz counterparts, and finally those with both 12.2~GHz counterparts and OH masers.



Therefore we suggest, similarly to \citet{Breen10a} but with a few additions, that 6.7~GHz methanol masers devoid of 12.2~GHz or OH maser emission present the youngest sources, followed by those with associated 12.2~GHz masers and finally those sources which exhibit both methanol maser transitions as well as OH maser emission. We further suggest that the methanol masers increase in both luminosity and velocity range with evolution and that this affect is greater in the 6.7~GHz sources.

\subsection{Association with GLIMPSE sources}\label{sect:glimpse}

The locations of the detected 12.2~GHz methanol masers, along with the 6.7~GHz sources with no detectable 12.2~GHz counterpart have been compared to both the GLIMPSE point source catalog and the catalog of Extended Green Objects (EGOs) presented in \citet{Cyg08}. The characteristics of infrared sources have been regarded as an excellent determinant of relative source ages over relatively long time-scales \citep[e.g.][]{Rath10}. The comparison between such a large number of methanol maser sources with the mid-infrared data is an important step in comparing the relative evolutionary stages associated with different maser species.


%

%

\subsubsection{Association with GLIMPSE point sources}

520 of our target 6.7~GHz methanol fall within either the GLIMPSE I or GLIMPSE II regions. Using an association threshold of 2 arcsec, 316 of our target methanol maser sources are coincident with a GLIMPSE point source listed in the catalog (a detection rate of 61~\%). The association rate for 6.7~GHz methanol masers without associated 12.2~GHz emission is higher (192 of 291, or 66~\%), than for sites exhibiting the two methanol maser transitions (127 of 229, or 55~\%). Methanol maser sources coincident with OH masers have a similar association rate as for the sources showing emission at 12.2~GHz, a rate of 57~\% (75 of 132). The similarity of the detection rates for sources with 12.2~GHz methanol masers or OH masers is expected given the high rate of coincidence for 12.2~GHz masers towards sites of 6.7~GHz methanol and OH masers.

\citet{Ellingsen06} found a GLIMPSE point source coincidence rate of 56~\% (23 of 41) for a subset of their 6.7~GHz sources that were detected in a complete search, and this rate dropped to 52~\% (29 of 56) when the full sample was considered. These association rate statistics, considered in conjunction with our own, suggests that the relatively more evolved star formation regions have fewer associations with GLIMPSE point sources. Evidence for this includes the fact that our sample of 6.7~GHz methanol masers yield the highest association rate (thereby including a high number of weak, possibly younger, sources) and this rate gets progressively less when stronger 6.7~GHz methanol masers or those associated with 12.2~GHz or OH maser emission, are considered. Since it is more likely for the more evolved sources to no longer be a point source in the GLIMPSE bands a lower association rate is expected.

\citet{Gallaway10} have undertaken extensive analysis of a near complete sample of methanol masers with associated GLIMPSE sources. Ordinarily analyses such as these have been limited to sources coincident with GLIMPSE point sources and to those IRAC bands with measured fluxes (which limits the sample). \citet{Gallaway10} have used an adaptive non-circular aperture photometry technique to measure the fluxes of GLIMPSE sources towards 512 methanol masers. From this large sample of comparisons, they show that the results of the smaller sample studies by \citet{Ellingsen06} remain valid.

\begin{figure*}
\includegraphics[scale=.4]{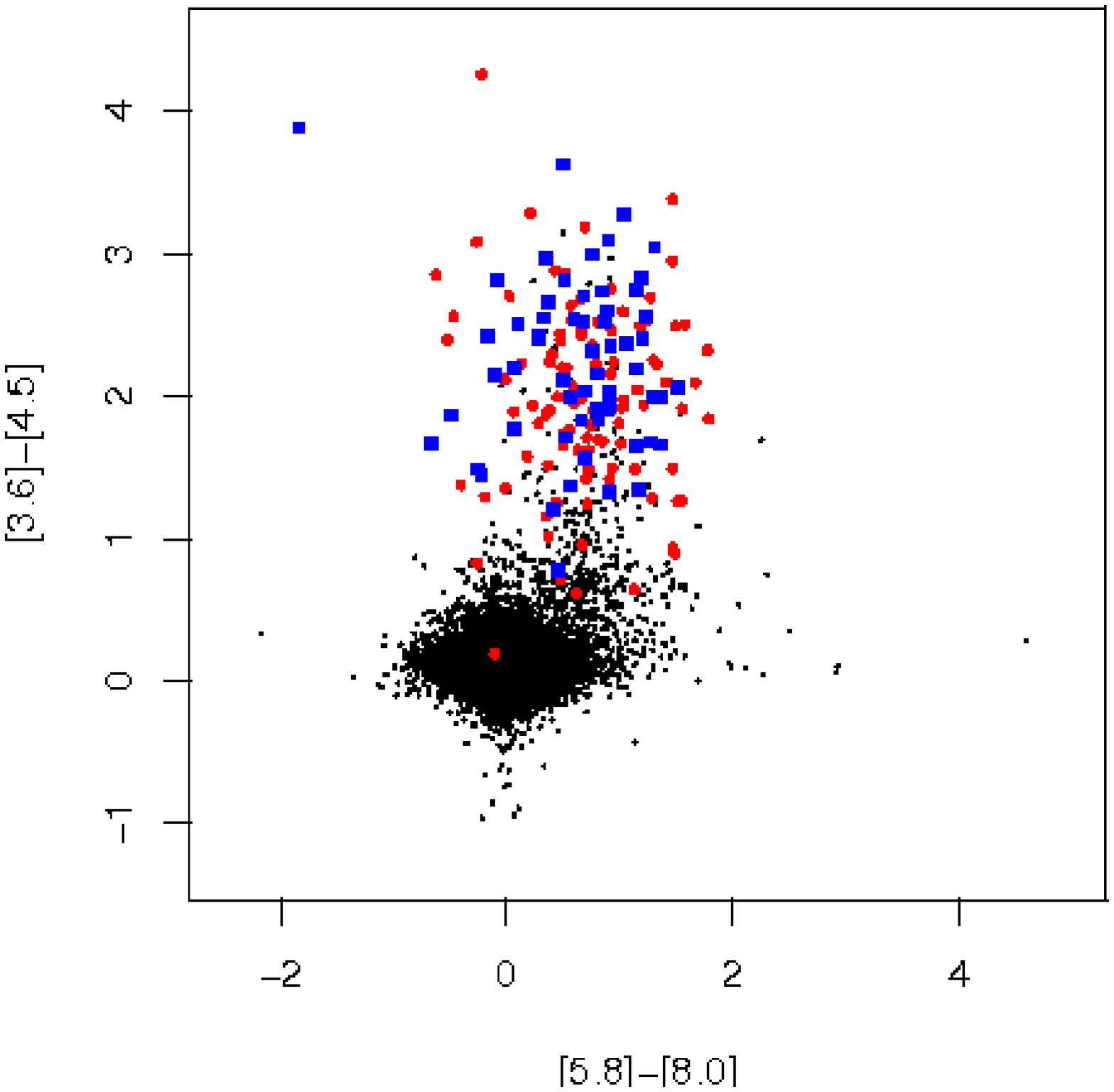}
\includegraphics[scale=.4]{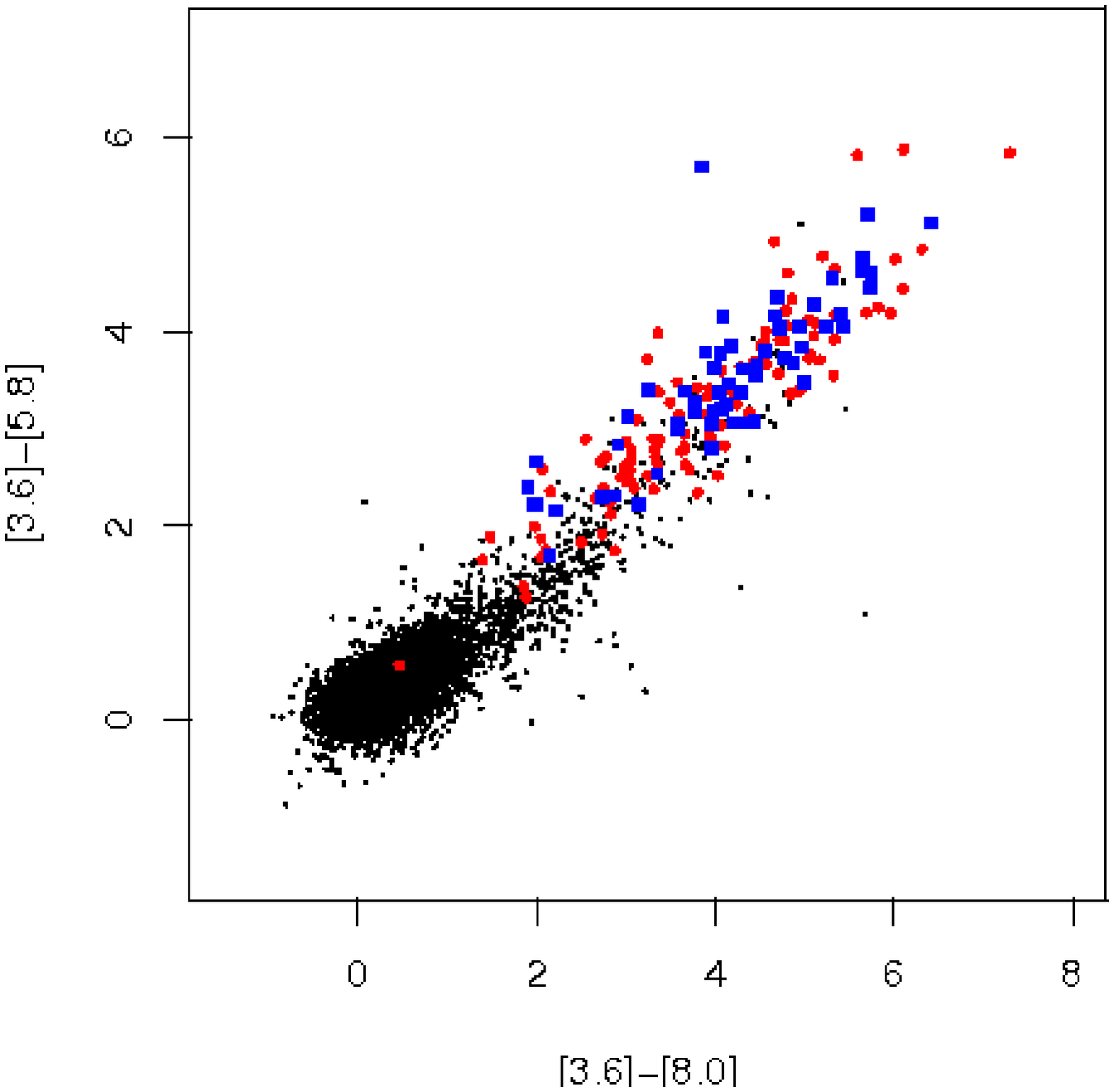}
\includegraphics[scale=.4]{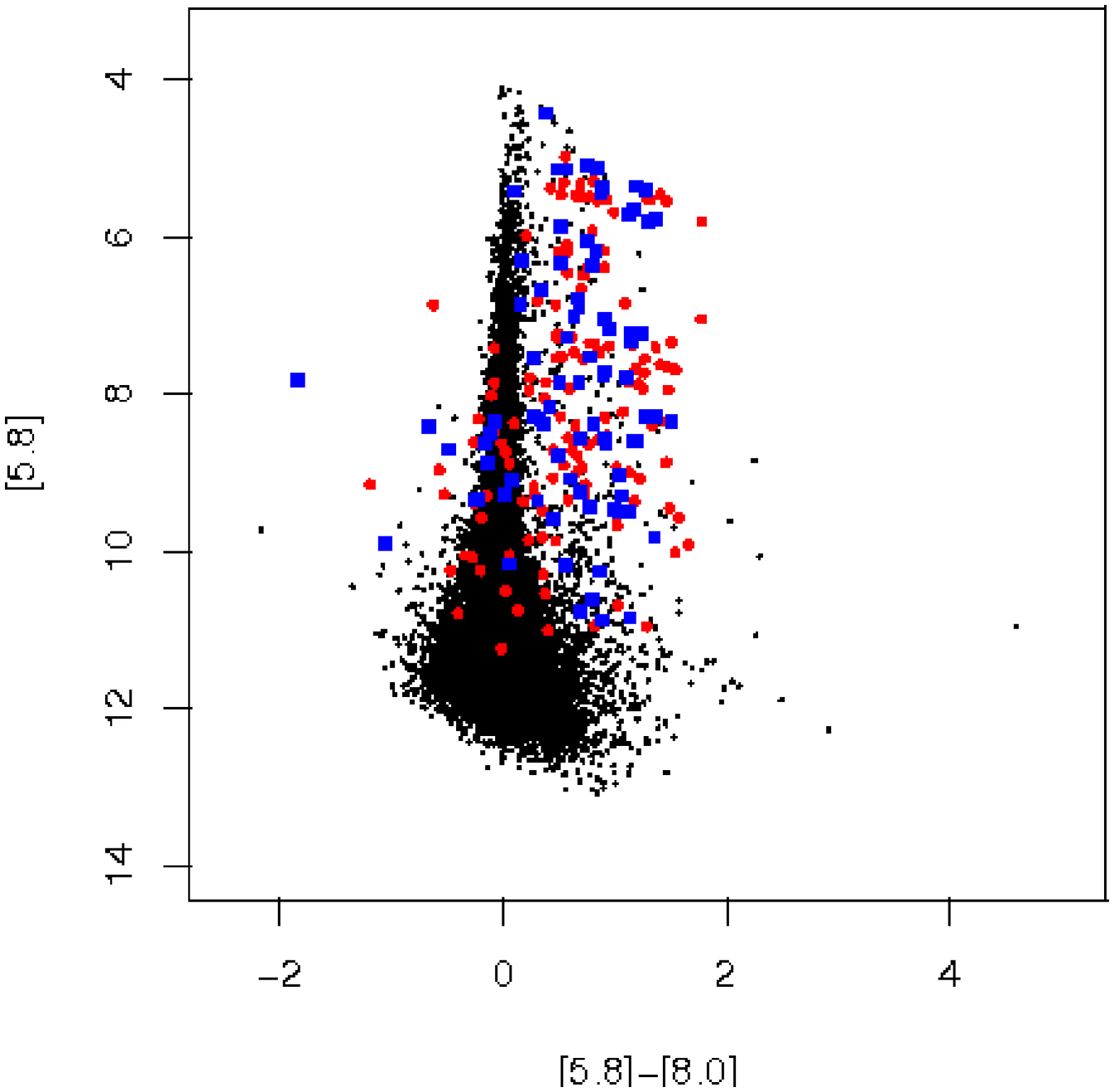}
\includegraphics[scale=.4]{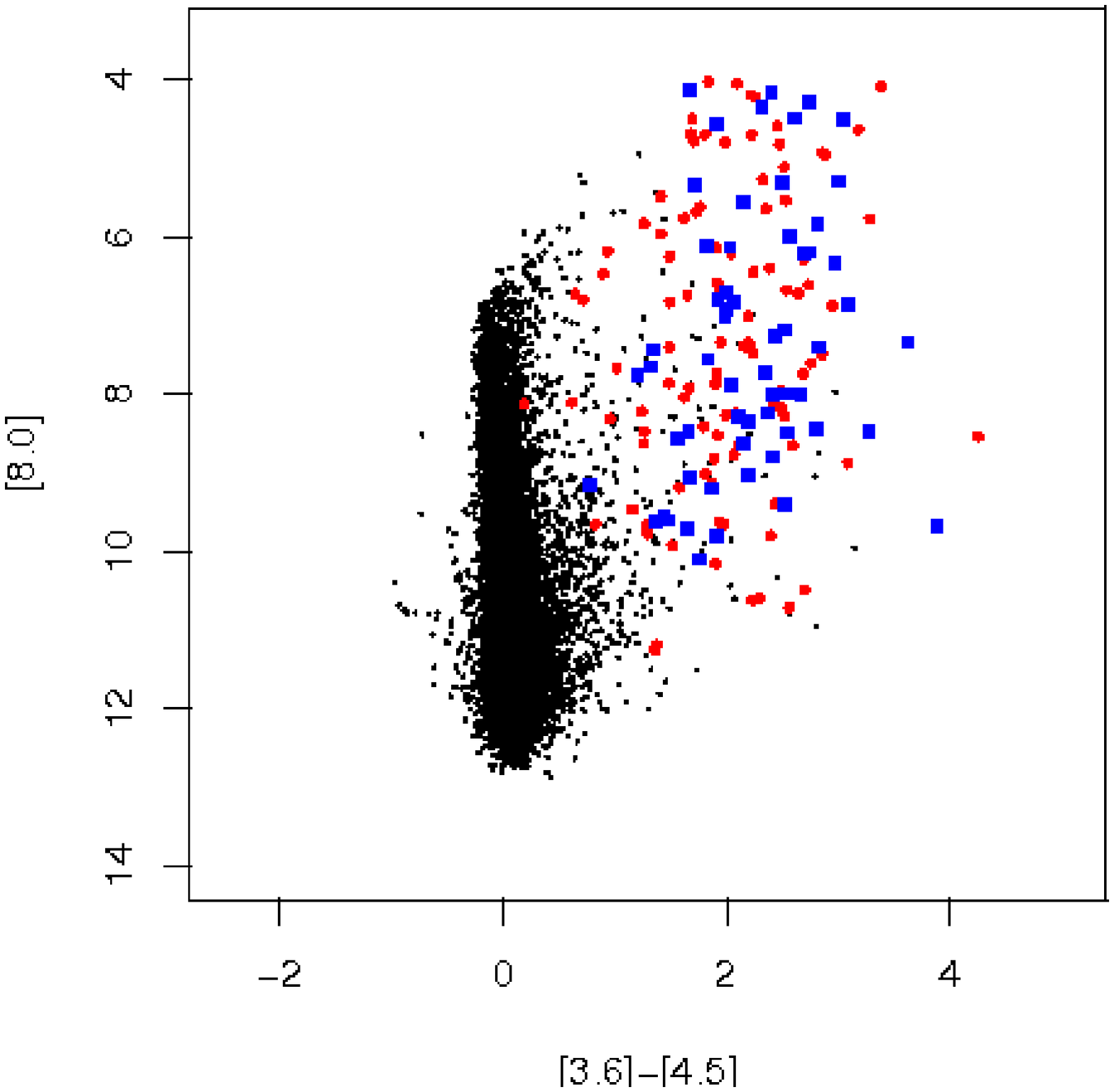}
\caption{GLIMPSE colour-colour and colour-magnitude diagrams. Clockwise from top left: [5.8]--[8.0] versus [3.6]--[4.5] colour-colour diagram; [3.6]--[8.0] versus [3.6]--[5.8] colour-colour diagram; [3.6]--[4.5] versus 8.0 $\mu$m colour-magnitude diagram; and [5.8]--[8.0] versus 5.8 $\mu$m colour-magnitude diagram. GLIMPSE point sources associated with sources showing emission at both the 6.7 and 12.2~GHz methanol maser transitions are represented by blue squares and those associated with only the 6.7~GHz transition are represented by red circles. In each plot the black dots represent all of the GLIMPSE point sources within 30 arcmin of $l$ = 326.5$^{\circ}$, $b$ = 0$^{\circ}$.}
\label{fig:glimpse}
\end{figure*}

Fig.~\ref{fig:glimpse} shows a sample of GLIMPSE colour-colour and colour-magnitude diagrams for the 6.7~GHz methanol masers coincident with GLIMPSE point sources. We find, similarly to \citet{Ellingsen06}, that the maser associated sources are significantly redder than the majority of the comparison sources (all of the GLIMPSE point sources within 30 arcmin of $l$ = 326.5$^{\circ}$, $b$ = 0$^{\circ}$) on the plot, lying in significantly different regions in each case. Evident from the [5.8]--[8.0] versus [3.6]--[4.5] colour-colour plot, the methanol maser associated sources appear to have a large relative excess of 4.5 $\mu$m flux. Interestingly, this plot along with the [3.6]--[4.5] versus 8.0 $\mu$m colour-magnitude diagram show that the majority of the sources that have similar 3.6 and 4.5 $\mu$m magnitudes are without associated 12.2~GHz emission. Aside from this, there are few obvious differences between the GLIMPSE point sources associated with both 6.7 and 12.2~GHz methanol maser emission and those associated with only 6.7~GHz methanol maser emission.

Additional plots similar to those presented in Fig.~\ref{fig:glimpse} were constructed with the methanol masers sources associated with OH maser emission distinguished from those with no associated OH maser emission. The sources with associated OH masers are relatively brighter in the 4.5 and 8.0 $\mu$m bands than those 6.7~GHz methanol masers with no OH maser emission. \citet{Ellingsen06} similarly noted that 6.7~GHz methanol masers with OH maser counterparts tended to have brighter 8.0 $\mu$m magnitudes. 

\citet{Ellingsen06} commented that the majority of masers associated with GLIMPSE point sources meet the criteria [3.6]--[4.5] $>$ 1.3 and 8.0 $\mu$m mag~$<$~10 as well as [3.6]--[8.0] $>$ 4.0 mag. From Fig.~\ref{fig:glimpse} it can be seen that these criteria are met by most of our methanol maser sources, although perhaps fail to capture some of the methanol maser sources, particularly a number of sources with no 12.2~GHz  maser counterpart. From our investigation it seems that these criterion could be refined to [3.6]--[4.5] $>$ 0.8,  8.0 $\mu$m mag $<$ 10 and [3.6]--[8.0] $>$ 2.0 mag. This may imply that a number of the younger sources have a less excessive 4.5 to 3.6 $\mu$m flux and likewise for 8.0 $\mu$m band.

Analysis of the GLIMPSE properties associated with the somewhat smaller sample presented in \citet{Breen10a} resulted in the conclusion that the masers themselves are much more sensitive to evolutionary changes than the associated mid-infrared emission. Apart from the few trends mentioned above, we similarly draw this conclusion.

\subsubsection{Association with Extended Green Objects (EGOs)}

The locations of the 6.7~GHz methanol maser sources were compared with the positions of EGOs presented in \citet{Cyg08}. We have adopted an association threshold of 10 arcsec to ensure that all associations were captured while minimising the number of chance associations. We find that 97 of the 580 6.7~GHz MMB methanol masers are associated with EGOs, and of these sources 49 have 12.2~GHz methanol counterpart. This equates to an overall 12.2~GHz detection rate of 50.5~\% (49 of 97) for 6.7~GHz methanol masers associated with EGOs and 42.2~\% (121 of 287) towards those without (after discarding 6.7~GHz sources that lie outside the regions inspected by \citet{Cyg08} for EGOs; 10$^{\circ}$ $<$l$<$ 65$^{\circ}$ and 295$^{\circ}$ $<l<$ 350$^{\circ}$, $b=$$\pm$1$^{\circ}$). This result means that in the case of a 6.7~GHz methanol maser that is associated with an EGO, the detection of a 12.2~GHz methanol maser is more likely than towards a 6.7~GHz methanol maser that is not associated with an EGO. The 12.2~GHz detection rates towards EGOs is statistically significantly higher than the detection rate for the entire 6.7~GHz sample which is 43~\%.

These comparisons also show that when taking into account only those 6.7~GHz methanol masers that fall within the regions inspected by \citet{Cyg08} that 25.3~\% (97 of 384) of 6.7~GHz methanol masers have associated EGOs while this increases to 28.8~\% (49 of 170) when you take into account only those 6.7~GHz masers that have associated 12.2~GHz methanol masers. \citet{Cyg09} found that the detection rate of 6.7~GHz methanol masers towards was much higher at a rate of $\sim$64 \% and suggested that this detection rate was higher than targeted searches that had used other selection criteria. While the detection rate is high, it is clear that a 6.7~GHz methanol maser search exclusively targeting EGOs may only recover approximately one-quarter of the total population.

69 of the 6.7~GHz methanol masers that are associated with EGOs have been searched for OH masers by \citet{C98}. 39 of these sources have an associated OH maser, an association rate of 57~\%. In the case where an EGO is associated with both a methanol and an OH maser there is a 74~\% chance that there will also be a 12.2~GHz methanol maser, this 12.2~GHz detection rate falls to 43~\% for EGOs with associated 6.7~GHz methanol masers but without OH masers. In comparison, comparing the locations of EGOs with water masers \citet{Breen10b} has shown that 35 of the 89 water-methanol-OH sources that fell within the regions \citet{Cyg08} inspected for the presence of EGOs, equating to an association rate of 39~\%. From these statistics it was concluded that EGOs persist into the stage of star formation that is evolved enough to have produced an OH maser, but not far past the stage where the developing \UCHII region has caused the methanol maser emission to cease. Our results support this hypothesis along with the idea that 12.2~GHz methanol masers are not present until well after the onset of the associated 6.7~GHz methanol maser emission.

\section{Conclusion}

We present the 12.2~GHz methanol maser properties of the 250 12.2 GHz methanol masers that we detect towards a complete sample of 580 6.7~GHz methanol masers detected in the MMB survey. The detection rate of 43~\% is lower than that of other large surveys of comparable sensitivity but this discrepancy can be attributed to the biases introduced from the nature of the target sources in these searches.

As found in previous searches, both the velocity ranges and the flux densities of the target 6.7~GHz sources surpass that of their 12.2 GHz companion in almost all cases. 80 \% of the detected 12.2 GHz methanol maser peaks are coincident in velocity with the 6.7~GHz maser peak and the peaks of the remaining 12.2~GHz sources are usually associated with a close secondary feature. There are a couple of instances where the peak of one or more of the 12.2~GHz methanol maser features surpasses that of the corresponding 6.7~GHz methanol maser peak. These unusual sources will be discussed in forthcoming catalog papers where we will explore in detail the specific data.

We find that there is a non-uniform 12.2~GHz methanol maser detection rate throughout the Galactic plane. In particular we find that the 12.2~GHz detection rate is highest in the longitude range 335 to 340$^{\circ}$. Furthermore the longitude range 270 to 305$^{\circ}$ has a lower 12.2~GHz methanol maser detection rate than the 315 to 350$^{\circ}$ range.

Some evidence of clustering of sources exhibiting 12.2~GHz masers (and likewise those with no 12.2~GHz emission) is apparent. We suggest that clustering may be a result of local bursts of star formation that have resulted in a number of nearby sources being at similar evolutionary stages.

Our analysis supports the evolutionary scenario present in \citet{Breen10a}, whereby the 12.2~GHz sources are associated with a somewhat later evolutionary stage than the 6.7~GHz sources devoid of this transition. Furthermore we find that both 6.7~GHz and 12.2~GHz methanol sources increase in luminosity as they evolve albeit more rapidly for the 6.7~GHz transition. In addition to this, evidence for an increase in velocity range as the sources evolve, is presented, and this also occurs at a greater rate for the 6.7~GHz sources. This implies that it is not only the luminosity, but also the volume of gas conducive to the different maser transitions, that increases as the sources evolve.

Comparison with GLIMPSE sources has revealed a similar coincidence rate between the locations of the masers and sources presented in the GLIMPSE point source catalog. It is evident that the GLIMPSE point sources that are associated with methanol masers have a large relative excess of 4.5 $\mu$m flux. We find little difference between the point source properties associated with the different maser categories (i.e. solitary 6.7~GHz sources, those with 12.2~GHz counterparts, and those associated with OH masers). There is a higher 12.2~GHz detection rate towards those 6.7~GHz methanol masers that are coincident with EGOs.\\
\\
\indent The Parkes radio telescope is part of the Australia Telescope National Facility which is funded by the Commonwealth of Australia for operation as a National Facility managed by CSIRO. This research has made use of NASA's Astrophysics Data System Abstract Service, the NASA/IPAC Infrared Science Archive (which is operated by the Jet Propulsion Laboratory, California Institute of Technology, under contract with the National Aeronautics and Space Administration) and data products from the GLIMPSE survey, which is a legacy science programme of the {\em Spitzer Space Telescope}, funded by the National Aeronautics and Space Administration.\\
\indent \facility {{\em Facilities:}} Parkes ()

\end{document}